\documentclass[fleqn,usenatbib]{mnras}

\usepackage{newtxtext}
\usepackage[frenchmath,varg]{newtxmath}

\usepackage[T1]{fontenc}
\usepackage{ae,aecompl}
\usepackage{graphicx}	
\usepackage{amsmath}	
\usepackage{multicol}
\usepackage{multirow}
\usepackage{booktabs}
\usepackage{textcomp}
\usepackage{url}
\usepackage{soul}

\newcommand{\halpha}{H$\alpha+$[\textsc{Nii}]\ }
	
\newcommand{\oiii}{[\textsc{Oiii}]\ }
\newcommand{\av}{$A_{\rm{V}}$}

\title[Central parsec of unobscured AGN]{Dust in the central parsecs of unobscured AGN: more challenges to the torus\thanks{Based on VLT programs 70.C-0565, 71.B-0292, 074.B-0404, 084.B-0568, 076.B-0493(B), 60.A-9026(A)}}

\author[]{
M. Almudena Prieto$^{1,2}$\thanks{E-mail: aprieto@iac.es}, 
Jakub Nadolny$^{1,2}$,
Juan A. Fern\'andez-Ontiveros$^{5}$,
Mar Mezcua$^{3,4}$
\\
$^{1}$Instituto de Astrof\'isica de Canarias (IAC), E-38205 La Laguna, Tenerife, Spain\\
$^{2}$Universidad de La Laguna, Dept. Astrof\'isica, E-38206 La Laguna, Tenerife, Spain\\
$^{3}$ Institute of Space Sciences (ICE, CSIC), Campus UAB, Carrer de Magrans, E--08193 Barcelona, Spain\\
$^{4}$Institut d'Estudis Espacials de Catalunya (IEEC), Carrer Gran Capit\`a, E-08034 Barcelona, Spain\\
$^{5}$Istituto di Astrofisica e Planetologia Spaziali, INAF--IAPS, Via Fosso del Cavaliere 100, I--00133 Roma, Italy
}
\date{}

\pubyear{2021}

\begin{document}
\label{firstpage}
\pagerange{\pageref{firstpage}--\pageref{lastpage}}
\maketitle

\begin{abstract}
A parsec-scale dusty torus  is thought to be the cause of  Active Galactic Nuclei (AGN) dichotomy in the 1/ 2 types, narrow / broad emission lines. In a previous work, on the basis of parsec-scale resolution infrared / optical dust maps it was found that dust filaments, few parsecs wide, several hundred parsecs long, were ubiquitous features crossing the centre of type 2 AGN, their optical thickness being sufficient to fully obscure the optical nucleus. This work presents the complementary view for type 1 and intermediate-type AGN. The same type of narrow, collimated,  dust filaments are equally found at the centre  of these AGN. The difference now resides in their location with respect to the nucleus, next to it but not crossing it, as it is the case in type 2,  and  their reduced optical thickness towards the centre, $A_V \lesssim 1.5\, \rm{mag}$, insufficient to obscure at  UV nucleus wavelengths. It is concluded that large scale, hundred pc to kpc long, dust filaments and lanes, reminiscent of those seen in the Milky Way, are a common ingredient to the central parsec of galaxies. Their optical thickness changes along their structure, in type 2 reaching optical depths  high enough  to obscure the nucleus  in full. Their  location with respect to the nucleus  and increasing gradient  in optical depth towards the centre could naturally lead to the canonical type 1 / 2 AGN classification,  making these filaments to play the role of the torus.  Dust filaments and lanes show equivalent morphologies in molecular gas. Available gas kinematic indicates mass inflows at rates  $ < ~ 1 M\odot~ yr^{-1}$.

\end{abstract}



\section{Introduction}\label{sec:intro}

In the Unified Model for Active Galactic Nuclei (AGN), the dichotomy between type 1 AGN --\,i.e. visible optical nuclei presenting optical broad-emission lines\,-- and type 2 --\,obscured optical nuclei without broad-emission lines\,-- is explained by the presence of a compact nuclear dust structure with a donut-like shape, the torus. This has with a height smaller than its radius and typical  size of a few parsecs \citep{Pier1992,Pier1993}. The orientation of this structure with respect to the line of sight is predicted to be the cause of the AGN dichotomy, as well as of the often cone-like, or simply, collimated distribution of ionised gas that emerges from the centre of AGN. Still, current infrared (IR) interferometric observations on the nearest, best studied AGN (\citealt{Kishimoto2011}; \citealt{Burtscher2013}; \citealt{Pfuhl2020})   at the relevant range  of $ 2\, \rm{\micron} - 10\, \rm{\micron}$ where the  torus  peak emission is expected are unveiling controversial results somewhat at odds with the prevailing scenario. In some cases the bulk of the parsec-scale dust is placed in the polar direction rather than in the plane of the putative torus (e.g. \citealt{Hoenig2013} for NGC\,3783 and NGC\,424; \citealt{Tristram2014} for Circinus; \citealt{LopezGonzaga2014} for NGC\,1068);  \citealt{LopezGonzaga2016}; \citealt{Asmus2019} and ref. therein, for several  AGN  imaged with single dish VLT / VISIR). In high obscured  sources as e.g. Centaurus\,A or in low luminosity AGN, e.g. NGC\,1052, M87,  the torus remains elusive down to sub-parsec scales (\citealt{Meisenheimer2007};  \citealt{Burtscher2010}; \citealt{Prieto2016}; \citealt{JA2019}). { In reference cases as NGC 1068, the extreme angular resolution provided by VLTI / GRAVITY at $2\, \rm{\micron}$ reveals a   ring like morphology of dust clouds and  step continuum, both inconsistent with  the expectation of  a geometrically thick clumpy torus (\citealt{Pfuhl2020})}. Nor the  recent GRAVITY results  in NGC 3783  point to the presence of a torus reconciled with a torus expectation. In this case,  a single  point-like continuum source and a  detached  dust cloud at scale of 0.14 pc  are uncovered (\citealt{Gravity2021}).
 
Nuclear obscuration may not necessarily be caused by a central, few parsec scale, structure. Dust filaments and lanes in the host galaxy can cause nuclear obscuration if they happen to cross the nucleus (e.g. \citealt{Malkan1998}; \citealt{Goulding_2012}; \citealt{Prieto2014}; \citealt{Buchner2017_II}), as well as circumnuclear dust structures in the form of disks or nuclear spirals, which may produce   obscuration and also be  a natural feeding mechanism of the nucleus (e.g. \citealt{PoggeMartini2002}, \citealt{Prieto2005}, \citealt{Prieto2019}). Nuclear filament and lanes are found to produce the often collimation of nuclear ionised gas in the narrow-line region (NLR; \citealt{Prieto2014}; \citealt{Mar2016}). The latter authors  show that the morphology of the \oiii$\lambda 5007$ ionisation cone in the classical type 2 Circinus Galaxy is caused by a pair of  dust filaments, few tens of parsec length,  that  converge into the nucleus defining a "V" shape within which the ionised gas expands. The role of the host galaxy as a potential source of nuclear obscuration is shown by \cite{Buchner2017_II}, who measure typical column densities for galaxy discs in the $N_{\rm H} \sim 10^{20}$--$10^{23}\, \rm{cm^{-2}}$ range and suggest that circa $40\%$ of the Compton-thin absorber might be caused by this gas in the disc. This would imply that a large fraction of the type 2 AGN population would be obscured by the host galaxy, regardless of the nuclear dust distribution and the orientation of the central engine.

Within the PARSEC project\footnote{\url{http://research.iac.es/proyecto/parsec}}, we are studying the effect of nuclear obscuration other than that associated to a torus in some of the nearest and best studied type 2 AGN, including cases such as NGC\,1068, Circinus, etc,  by means of parsecs-scale resolution dust maps constructed from Adaptive-Optics in the near-IR and \textit{HST} optical images at comparable angular resolution (\citealt{Prieto2014}; \citealt{Mar2016}). We found that dust filaments and lanes are ubiquitous to the centre of all the type 2 nuclei analysed, their location being  always crossing the nucleus in all the galaxies, the attenuation produced by these dust structures being in all cases sufficient to hide these nuclei at optical wavelengths. Furthermore, the morphology of the nuclear --\,optical\,-- ionised gas is in all cases profiled by the shape and location of these dust filaments and lanes. These structures play the role of the torus but extend from tens of parsec to kpc into the host galaxy. As being part of the host galaxy, these dust structures are expected to be also ubiquitous to type 1 AGN, presenting similar morphology and location. This paper examines their existence in an equivalent sample of the nearest type 1 and intermediate-type  AGN and evaluates the obscuration they may introduced on these unobscured nuclei. The aim of this work is to uncover the location and morphology of obscuring material in the central kpc of the galaxies sample, and its distribution and location with respect to that of central ionised gas, and  of the nucleus.

The paper is organised as follows: Section\,\ref{sec:data_source} gives a short description of the data, the image alignment process, and the construction of  colour and extinction maps; Section\,\ref{sec:results} identifies the galaxies nucleus location and discusses the nuclear dust and ionised gas morphologies; Section\,\ref{sec:conclusions} discusses the results in the context of the unified models and our previous studies of type 2 nuclei.

\section{Observations and Analysis}\label{sec:data_source}

This work is based on a high-angular-resolution, FWHM$ < 200\, \rm{mas}$, imaging database of a sample of the nearest AGN of type 1 to 1.9 (see Table \ref{tab:general_info}). The dataset spans the optical to near-IR range. The near-IR data are collected with the Very large Telescope (VLT) using Adaptive Optics (AO) assisted instrumentation, the optical/UV data are taken with \textit{HST}.

\subsection{Data acquisition and reduction}\label{sec:data_reduction}
The near-IR images are taken with the { NaCo adaptive-optics (AO) assisted camera (Nasmyth Adaptive Optics System - Coude Near Infrared Camera)} at the VLT. The nuclei of all the galaxies are { bright enough to be used as AO guide star}. This has the advantage of maximising the angular resolution delivered by the AO system. Imaging in the \textit{Ks} band ($\lambda_c = 2.18\, \rm{\micron}$, $\Delta\lambda = 0.35\, \rm{\micron}$) and the \textit{H} band ($\lambda_c = 1.6\, \rm{\micron}$, $\Delta\lambda = 0.33\, \rm{\micron}$) is used. The field of view (FoV; $28'' \times 28''$) of NaCo covers the central few kpc of each galaxy. The average resolution of the dataset is FWHM\,$\sim 0\farcs16$, measured on point-like sources in each image (Table \ref{tab:general_info}, col.\,5). All the images were taken in a dither mode and reduced using the ESO pipeline Eclipse (\citealt{Devillard1999}). { The use of this  package is preferred to the standard ESOreflex due to our extensive knowledge of the package and the possibility to modify  parameters for background subtraction and frames registration optimization}. Data reduction includes background subtraction, registration of single exposure frames and final image combination.

The optical images are taken from the Wide-Field Planetary Camera 2 (WFPC2) and the Advanced Camera for Surveys (ACS) of the \textit{HST}. The dataset includes broad-band images in \textit{B}, \textit{V} and \textit{I} filters, namely \textit{HST} filters F438W, F475W, F547M, F550M, F555M, F28X50LP, and narrow-band filters centred on \halpha and \oiii emission lines, i.e. \textit{HST} filters FR533N, F658N. Data reduction followed the standard \textit{HST} pipeline procedure for dithered images which is to combine the images and remove cosmic rays. To improve the astrometry precision of the images, which is essential for this work, the \textsc{MultiDrizzle} PyRAF-based package \citep{Koekemoer2003} was used to correct for geometric distortion.
 
\begin{figure*}
	\includegraphics[width=0.9\textwidth]{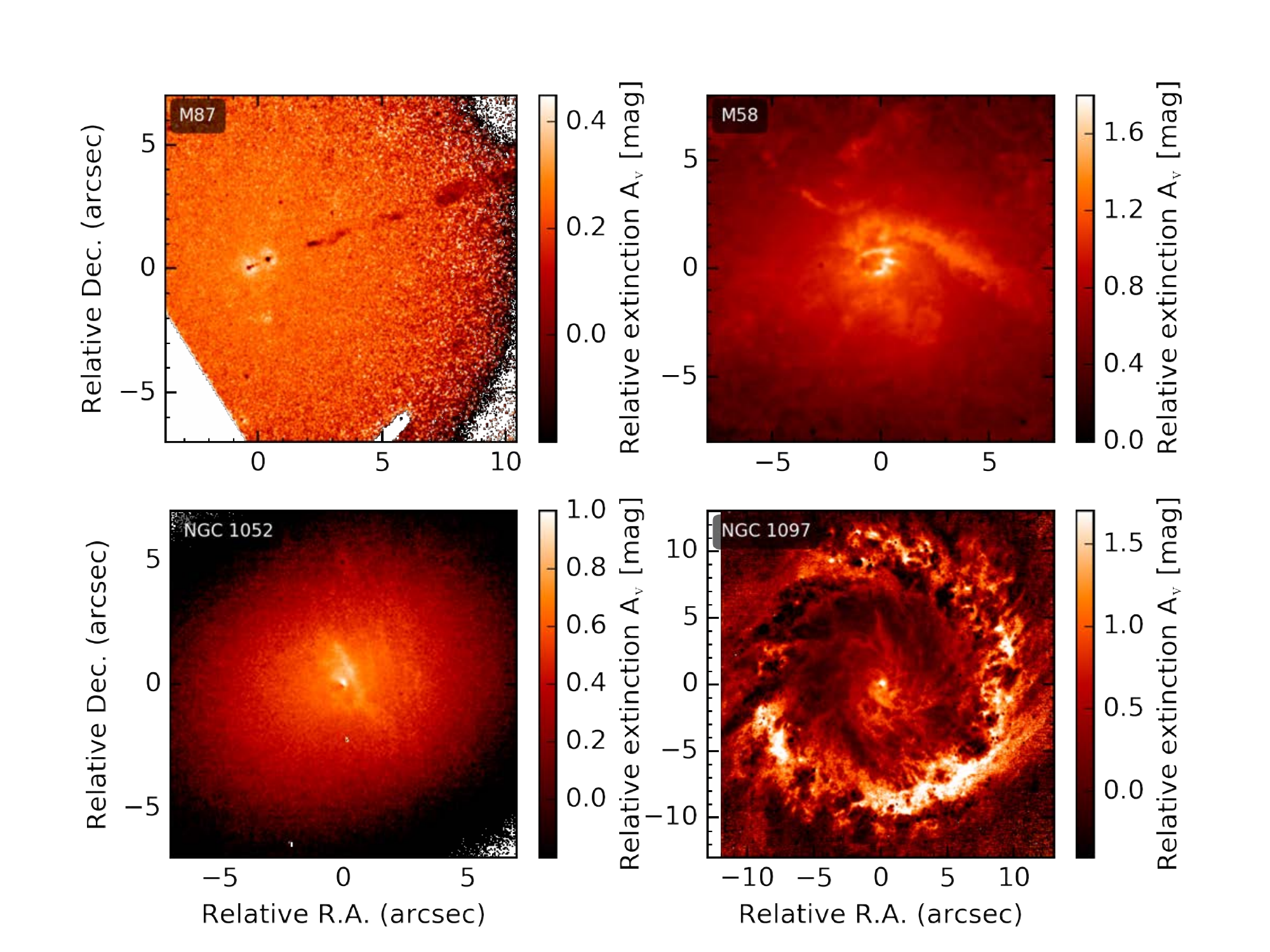}
  \caption{Extinction maps of M87, M58, NGC\,1052, NGC\,1097.}\label{fig:ext_1st}
\end{figure*}

\subsection{Image registration}\label{sec:ImgReg}
The comparable resolution of the VLT-AO images in the IR to  that of  the HST in the optical-UV  allows us to produce an accurate registration  of the whole set   of images at the mas level. This accuracy allow us to locate  precisely the nucleus and any dust structure in the central parsec of the galaxies sample.

The procedure for image registration was developed in \citet{Fernandez2009}, and  further applied in  e.g. \citealt{Prieto2014}; \citealt{Mar2015}; \citealt{Mar2016}. Briefly, per each galaxy, the set of optical and IR images were aligned to a common reference frame, this being defined by the VLT-NaCo \textit{Ks} or \textit{H} band images. To that aim, all the images were examined for several point-like sources, typically compact 
(FWHM $\lesssim 5\, \rm{pc}$) star forming clusters other than the galaxy nucleus, common to the FoV of all of them. These sources were used for image registration using a minimum square fit, the IRAF\footnote{\url{www.iraf.noao.edu}} \textsc{imalign} task was used. We also used our own scripts to cross-check the IRAF results. Image shifts were derived from the offsets found among the mean central position of the point-like sources at each filter image. The positional errors range between $10$ and $40\, \rm{mas}$ (Table\,\ref{tab:general_info}, column 9), and are measured as the standard deviation of the offsets found per each point-like position and filter. Our precision is at pixel or subpixel level in most cases (Table\,\ref{tab:general_info}). { Positional errors depend on the quality of the Gaussian fit to determine the source's centroid. This in turn depends on source signal to noise and  circularity - sources at the edges of the images are more affected by image distortion, and anisoplanetism in the case of the AO images.}

Figures\,\ref{fig:m87}--\ref{fig:ngc7469} (Appendix \ref{sec:figures}) present per each galaxy the images used in the work. Top left panels show the optical \textit{HST} image with the IR (\textit{Ks} or \textit{H} band) on top in contours. The blue circles mark the point-like sources, seen in both the optical and IR images, used for the alignment process.  By default the nucleus is not used in the alignment process, with  the exceptions of NGC\,3081 and NGC\,3783 (Table \ref{tab:general_info}) with no point-like sources other than the nucleus in the field of view. In these cases the registration relies on the nucleus only.

\subsection{Dust maps and extinction in the central kpc}\label{sec:DustMaps_extinction}
To uncover the dust location and morphology, colour maps of the central kpc of the galaxies in the sample were constructed as a ratio of an IR to an optical image, usually VLT-NaCo \textit{Ks} band to \textit{HST} \textit{V} band. These colour images are effectively dust maps: regions suppressed by dust in the optical/UV appear enhanced in the IR-to-optical ratio. They are shown in Figs.\,\ref{fig:m87}--\ref{fig:ngc7469}. In these dust maps,  the brightest regions pinpoint where the optical light is suppressed -  hence brighter in Ks-band - to some level, presumably by dust, the darker regions are either dust free or sites of star formation, or the nucleus itself. The nucleus being type 1 or intermediate-type  is often brighter in the optical than in the IR, hence its relative dark appearance relative to the dust filaments around  in the maps. 

On top of the colour maps, the IR image is shown in contours, presenting smooth isophotes up to the central parsecs due to its transparency to dust. It can be noticed that obscuring material, i.e. the brighter colours in the maps, is found in all the galaxies, and systematically at the central parsecs. The angular resolution of the IR images, of few tens of parsec (Table\,\ref{tab:general_info}), allows us to dissect the morphology of the obscuring material, this being mainly made of narrow filaments and lanes surrounding or even crossing the nucleus. Most of these structures extend up several hundred parsec to kpcs scale, the larger scales  set by the available field of view of the images. The only exception is M87, effectively free of dust over its central kpc.

An estimate of the extinction, \av, produced by this material is evaluated from these maps following the same procedure as that applied for the type 2 sample in \citet{Prieto2014} and \citet{Mar2015}. Briefly, for each galaxy, colours are measured at the locations of the dust filaments and lanes in regions as close as possible to the nucleus, and then compared with equivalent measurements in dust-free regions in the  in the neighbourhood selected by visual inspection. 
This  comparison  yields a measurement of the { relative extinction \av\ towards the centre,  typically as close as  a few ten of parsec to the AGN. }

In all cases, colours are measured in circular apertures with radius that match the smallest dimension of the dust structures, typically a few tens of parsec, i.e. in the  0.1  -- 0.3\arcsec radius range,  in line  with the angular resolution of the images.
The regions selection is a compromise between they being as close to the centre as possible and  avoid the AGN light contribution.   {The same aperture sizes are used for the dust-free regions. These  are selected at distances of tens parsecs from the centre, wherever they could be isolated from filaments and lanes. }

In most cases, it was possible to construct various K-band / optical colour maps depending on the availability of \textit{HST} continuum images. This allowed us  to check the reliability of the estimated $A_V$, values can be compared in Table\,\ref{tab:extinction}.  Several measurement at different regions along the filaments and  in the dust-free regions were produced, and their  average, value and typical location,   are given  in  Table 2 per each color map.
It should be noted that selected dust-free regions may still contain some dust, in which case the inferred \av\ is a lower limit. Still, the agreement in \av\ from independent colour maps is remarkable, within less than $0.5\, \rm{mag}$ in most cases.  

Also we note that the reported {\av\ values do not measure  any extinction at the nucleus itself but at its surrounding, typically within  ten of parsec  from the centre:} nuclear colors are not representative of extinction as  both non-thermal emission and presumably  dust in emission from the AGN also contribute to these colors.

\begin{figure*}
\includegraphics[width=0.9\textwidth]{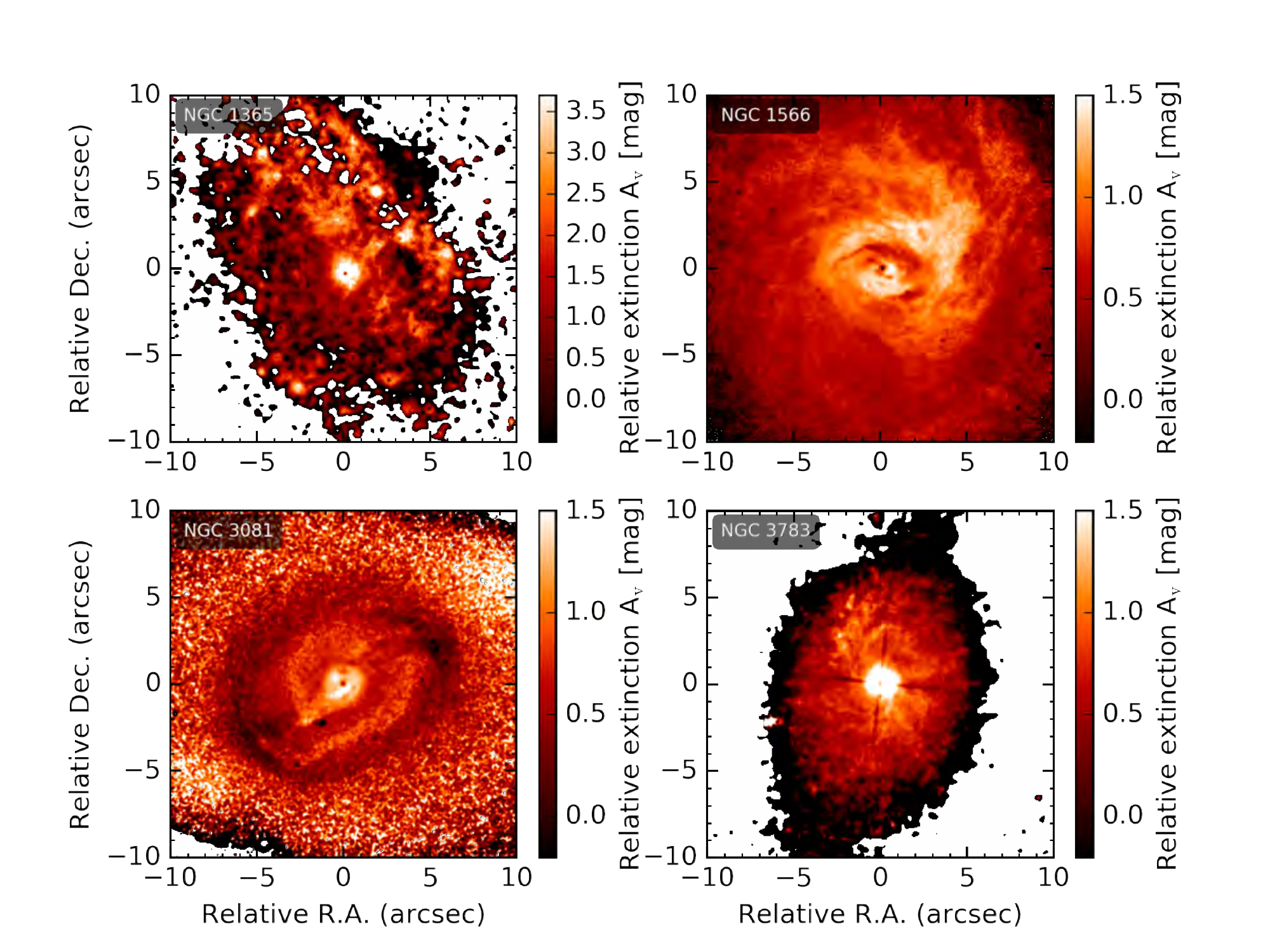}
\caption{Extinction maps of NGC\,1365, NGC\,1566, NGC\,3081, NGC\,3783.}\label{fig:ext_2nd}
\end{figure*}

Figures\,\ref{fig:ext_1st}, \ref{fig:ext_2nd} and \ref{fig:ext_3rd} show the extinction maps for the central kpc of all the galaxies. The extinction maps are based on the colour maps that best illustrates the dust absorption in the galaxies, namely \textit{V} (or closer to \textit{V} band) $-$ \textit{Ks} colour, and converted to an extinction map as above described. Note that some high values in some extinction maps refer to regions where the optical emission is by nature stronger than in the IR, e.g., the young star forming regions in NGC\,7469, NGC\,1097, NGC\,1365,     jet  in M87, or in the  step UV continuum of the  nucleus itself in most galaxies. At these locations, included the nucleus as just discussed above,  the maps do not measure dust extinction but simply the stellar or non-thermal contribution.

\subsection{Nuclear ionised gas}
\label{sec:ionisedGas}
Most of the AGN in the sample show nuclear extended ionised gas. This section investigates  the relative location of this ionised gas with respect to the nuclear filaments and lanes. The ionised gas is traced by \textit{HST} \halpha$\lambda 6548$, $\lambda 6563$, $\lambda 6583$ and / or \oiii$\lambda 5007$ line maps. The line maps are equally registered to mas accuracy to the rest of the continuum images dataset and thus, its location with respect to that of the dust can be strictly derived without any a priory assumption of its expected position or that of the nucleus in the dust maps (Section\,\ref{sec:DustMaps_extinction}).

Specifically, continuum-free line images were extracted from \textit{HST} narrow-band FR533N and F658N images centred in \oiii and H$\alpha$+[NII], respectively.
Prior to the subtraction, the \textit{HST} continuum images were re-scaled to account for the different filter bandwidth and transmission. The continuum image used is the closest in wavelength to the line emission (Section\,\ref{sec:data_reduction}). To estimate the continuum contribution, the continuum image was multiplied by a scaling factor chosen iteratively until the continuum-free line image did not show negative values. The resulting continuum-free line images are shown in contours on top of the dust-maps in Figs.\,\ref{fig:m87}--\ref{fig:ngc7469}.

\begin{table*} 
	\caption{\textbf{Description:}(1) Object; (2) distance in Mpc; (3) linear scale; (4) references for distances: [1] \citet{Blakeslee2009}; [2] \citet{Ruiz-Lapuente1996}; 
		[4] \citet{Tully1988}; [5] \citet{Kanbur2003}; [6] \citet{Sandage1994}; [7] NED -- adopting $H_0 = 73\, \rm{km\,s^{-1}\,Mpc^{-1}}$; [8] \citet{Genzel1995}; (5) averaged IR data resolution measured at point-like sources used in the alignment process; (6) filters used in this work; (7) ionised gas: \halpha$\lambda 6548$, $\lambda 6563$, $\lambda 6583$ and \oiii$\lambda 5007$; (8) number of point-like sources common in the FoV used for the alignment; (9) error averaged over the alignment process; (10) shift between IR and optical peak emission. For all  objects  but one, M58, no shifts are found within the registration error. \maltese\ indicates that the image registration was done only with the nucleus, thus the optical-IR shift cannot be determined.}\label{tab:general_info}
		\begin{tabular}{cccccccccc} 
			Object		& D 			&  Scale 	& Ref. 	&  Resolution 				& Filters 				&  Ionised 				&  Clusters 	& Errors		& Optical-IR  	\\
			& (Mpc) 	& ($\rm{pc/\arcsec}$)					&			& {IR band} (\arcsec)	&					&  gas 					&					& (mas) 	& {distance} (mas) \\ 
			(1)			&(2)			&(3)				&(4)		&	(5)				&(6)								&(7)						&(8) 			& (9)		&(10)\\ \hline
			M87 			& 16.7 	& 81 			&[1] 	&	0.12				&F475W, F658N, F814W, \textit{Ks}	&\halpha 				& 3				& 10			& -   	\\
			M58 			& 21			&102 			&[2]		&	0.16				&F330W, F547M,	F658N, \textit{Ks}	&\halpha 				& 3				& 10 		& 69 	\\
			NGC\,1052	&18			&87				&[3]		&	0.22				&F28X50LP, F658N, \textit{J}, \textit{Ks} & \halpha 			& 3				& 40			& -	\\
			NGC\,1097 &14.5		&70				&[4]		&	0.16				&F336W, F438W, F814W, F658N, \textit{Ks}	&\halpha				& 7				& 30			& -	\\
			NGC\,1365 & 18.5 	&90				&[5]  	&	0.12				&F336W,F555W, F814W, FR533N, \textit{Ks}	&[\textsc{Oiii}]	& 6 				& 20			& - 	\\ 
			NGC\,1566 &20			&96				&[6]		&	0.16				&F438W, F814W, F658N, \textit{J}, \textit{Ks}	&\halpha				& 3				& 20			& -	\\
			NGC\,3081	&33			&160			&[7]		&	0.4				&F547M, FR533N, \textit{Ks}	&[\textsc{Oiii}]	& -				& 10			& \maltese	\\
			NGC\,3783	&40.3		&195			&[7]		&	0.07				&F550M, \textit{J}, \textit{Ks}					&-							& -				& 20			& \maltese	\\ 
			NGC\,7469	& 66			&320			&[8]		&	0.11				&F550M, F814W, FR533N, \textit{H}	&[\textsc{Oiii}] 	& 4 				& 30 		& -  	
		\end{tabular}
	\end{table*}

\begin{table*} 
	\caption{\textbf{Description:} (1) Object; (2) AGN type: Sy -- Seyfert, L -- LINER; (3) References for column (2): [1] \citet{Veron2006}, [2] \citet{Ho1997b}; (4) Bolometric luminosities; (5) References for column (4): [1] \citet{Fernandez2012}, [2] \citet{Dudik2007}, [3] \citet{Jorsater1995}, [4] this work, [5] \citet{Prieto2010}; (6) Extinction $A_V$ towards the nucleus, measured at a radius given in col. (7), from various \textit{HST}--VLT-NaCo colour images; (7) and (8) Maximum distance to the nucleus of the reference aperture locations in the dusty and dust-free regions, respectively. \dag \ For NGC\,1097  two additional extinction values: $(U-I) \sim 1.6\, \rm{mag}$ and $(B-I) \sim 1.7\, \rm{mag}$ were produced - not shown. { The dust-free region  in the U-I map had to be measured at $\sim70'~ arcsecs$ from the nucleus.},}\label{tab:extinction}
		\begin{tabular}{ccccccccccccc}
			 Object 	& AGN type	& Ref.	& $L_\mathrm{\rm bol}$	&  Ref.	& \multicolumn{6}{c}{$A_{\rm{V}}$ (mag)} & \multicolumn{2}{c}{ Nuclear distance}\\ \cmidrule(r){6-11} \cmidrule(r){12-13}
			& 	&			& ($\rm{erg\,s^{-1}}$)							&		& $B-K$ & $V-K$ & $V-H$ & $V-I$ & $I-K$	& $J-K$	& Dust		& Dust free \\ \cmidrule(r){6-11}
			(1)			&(2)		&(3)				& (4)							&(5)	&\multicolumn{6}{c}{(6)}												& (7) 			&(8) 	\\ \hline
			M87 		&L1		&[1]				&1.2$\times10^{42}$		&[1]		& 0.4	& -			 &-	&-				&0.		&-				&1.0$''$		&1.5$''$  \\
			M58 		&L1.9	&[1]			 &3.2$\times10^{42}$	  &[2]		& -			&1.8		&-		&-				&-			&-			&0.5$''$	& 3.5$''$\\
			NGC\,1052 	&L1.9 	&[2]		 &9.3$\times10^{42}$	  &[1]		& -			&0.9 	    &-		&-				&-			&0.7		&0.5$''$	& 2$''$\\
			NGC\,1097$^{\dag}$&L1 &[1] 	&3.8$\times10^{41}$      &[5]		&1.8	    &1.2	   &-		&-				&1.7		&1		&0.5$''$	& 2.5$''$-11.6$'$  \\
			NGC\,1365	&Sy 1.8 &[1]  		 &1.6$\times10^{44}$	   &[3]		& -			&2			&-		&-				&3.5		&-			&1$''$		& 8$''$ \\ 
			NGC\,1566 	&Sy 1.5&[1]			 &5.5$\times10^{42}$	  &[5]		&1.4	   &-			& -		&1.2			&-			&1.2			&0.5$''$	& 4$''$-10$''$  \\
			NGC\,3081	&Sy 1 	&[1]		  &3.3$\times10^{41}$	    &[4]	 & -		  &1.4		&-		  &-				&-			&-			&1$''$		& 5$''$ \\ 
			NGC\,3783	&Sy 1.5&[1]	  		&1.8$\times10^{44}$	       &[5]		& -			&1.1		&-		 &-				&-			&0.5		&1.5$''$	& 4$''$ \\ 
			NGC\,7469 	&Sy 1.5&[1]	  		&2.5$\times10^{44}$	      &[5]		&-			&-			& 2	   	 &1.9			&-			&-			&0.5$''$	& 3$''$  \\
		\end{tabular}
	\end{table*}

\section{Results}\label{sec:results}

\subsection{Nuclear dust: morphology, location, and extinction}
The dust maps (Section\,\ref{sec:DustMaps_extinction}) show the ubiquity presence of obscuring material in the central parsecs of all the study galaxies, the sole exception is M87. This obscuring material resolves in long and narrow filaments and lanes. In some cases, the filaments are restricted to the central few hundred parsecs, e.g. NGC\,1052, $\sim 200\, \rm{pc}$ (Figs.\,\ref{fig:ext_1st} and \ref{fig:m58-uv}), in all others the obscuring material extends over the whole FoV of the maps, $\sim 1$--$2\, \rm{kpc}$ radius in average. The most conspicuous structures are the filaments running rather collimated from from kpcs distance all the way to the centre: they either follow an almost straight path to the centre, e.g. M58 (Fig.\,\ref{fig:ext_1st}), or circularise to form a nuclear spiral: e.g. NGC\,1097 (Fig.\,\ref{fig:ext_1st}, see also \citealt{Prieto2005}), NGC\,3783, NGC\,1566 (Fig.\,\ref{fig:ext_2nd}) or a combination of both, e.g. NGC\,3081 (Fig.\,\ref{fig:ext_2nd}). In NGC\,1365 (Fig.\,\ref{fig:ext_2nd}) and NGC\,7469 (Fig.\,\ref{fig:ext_3rd}), kpc-extent lanes and filaments cross the central region in multiple directions. Extinction maps of the central kpc region (Section\,\ref{sec:DustMaps_extinction}) indicate inhomogeneous \av\ across the region, the highest \av\ are found at the more conspicuous structures in the filaments and lanes, and overall, \av\ increases toward the centre along a lane or filament. 

These general properties apply to a sample of the nearest and brightest type 1, 1.5 to 1.9 AGN. An equivalent study in type 2 nuclei, in \citet{Prieto2014}; \citet{Mar2015} and \citet{Mar2016}, reveals the same characteristics in terms of ubiquity, morphology and scale-length of the central obscuring material. The sole difference is the inferred \av\ towards the nucleus, a factor 2 to 3 larger tin type 2 than in the present type 1--1.9 sample. The dust filaments that cross the nucleus in the type 2 AGN dim or even totally obscure
the nucleus at optical/UV wavelengths, and the usual bright, point-like $2\, \rm{\micron}$ source seen at the centre of all these nuclei, and identified with the AGN, appears always shifted with respect to the optical photometric peak, on scales of few parsecs \citep{Prieto2014}. Yet, in the present sample the optical and IR peak, amid the filaments and lanes crossing the centre, always coincide in position, within our precision astrometry of few tens of mas, i.e. parsec to sub-parsec scales (Table\,\ref{tab:extinction}).

In the present sample, the nucleus is a bright point-like source up to the UV at \textit{HST} resolutions: \cite{Prieto2005} for NGC\,3081; \cite{Maoz2007} and \cite{Fernandez2012} for NGC\,1052, M87; \cite{Prieto2010} for NGC\,1097, NGC\,1566, NGC\,3783, NGC\,7469; this work for NGC\,1365 and M58. A particular case is M58, type 1.9 class which does not show a clear point-like optical counterpart to the nuclear $2\, \rm{\micron}$ peak at wavelengths about $5000$\,\AA \ (\textit{HST} F547M) but it shows in the UV (\textit{HST} F330W).
The $5000$\,\AA \ peak is shifted by $\sim 70 \pm 10\, \rm{mas}$ ($\sim 7 \pm 1\, \rm{pc}$) in direction North-East with respect to the $2\, \rm{\micron}$ peak.
 In the UV, it is a strong point-like source coinciding with the $2\, \rm{\micron}$ peak within a precision of $10\, \rm{mas}$ (Fig.\,\ref{fig:m58}, Fig. A3;  Table\,\ref{tab:general_info}). M58 is a low luminosity AGN, $L_{\rm bol} \sim 10^{42}\, \rm{erg\,s^{-1}}$, and the nucleus is largely shadowed by the much stronger stellar light in the optical, not yet  in the UV where the stellar light usually drops.
 
Overall, given the filamentary and intricate morphology of the central obscuring material in type 1 as 2 alike, we have to conclude that in type 1 and intermediate-type AGN we are seeing these nuclei through either a rather dust-free window --\,the nuclear filaments just happen not to cross the line of sight\,-- or a relatively tenuous material which, depending on the nuclear brightness, is insufficient to obscure the nucleus at optical / UV wavelengths. For example, a nucleus with $L_{\rm bol} \sim 10^{44}\, \rm{erg\,s^{-1}}$ may remain partially visible behind an $A_V \sim 3\, \rm{mag}$ lane (e.g NGC\,1365), or be unaffected by $A_V \sim 1\, \rm{mag}$ filaments (e.g. NGC\,3783). These type of structures were reported by \cite{Malkan1998} after their \textit{HST} imaging survey of 200 Seyfert galaxies with the WFPC2 / F606W continuum filter. They found a preponderance of nuclear obscuring structures, with morphology and length as those find here, in the Seyfert type 2 group; in the present work, the high angular resolution of the IR / optical colour maps and the fact that colours rather than single images are more sensitive to dust, allow us to trace these structures down to the inner few parsec in type 2 and 1 alike, and regardless of activity level, from Seyfert to the low-luminosity AGN class.

\subsection{Nuclear ionised gas location relative to the dust}\label{sec:results_ionised_gas_and_dust_rel_dist}
In our equivalent study of type 2 AGN, an strict spatial anticorrelation between the nuclear extended H$\alpha$ or \oiii$\lambda 5007$ gas and the dust filaments and lanes surrounding the nucleus is found. Thus, it was concluded that the typical nuclear ionisation cones and collimated morphologies that characterise the narrow line region in the sample are not caused by a parsec-scale collimating torus as envisaged by the AGN unified model but simply by the location / morphology of the filaments and lanes in the central region (\citealt{Prieto2014}). Should that be the case, an equivalent finding in type 1 AGN may be apparent. The result is illustrated in Figs.\,\ref{fig:m87}--\ref{fig:ngc7469}. In these figures, per each galaxy,  the continuum subtracted \textit{HST} \halpha or \oiii image, depending on availability, is shown in contours  on top of the corresponding dust map - in all cases, the  first contour is $3 \sigma$ above the background.

It can be seen from the figures that the morphology of the ionised gas is for most cases  rather symmetric around the nucleus, except for the preferential location of the gas at one side of the nucleus in  NGC\,1365  or the somewhat  collimated morphology in  NGC\,3081,   North of the nucleus, or along the jet in  NGC\,1052.  As the nucleus is always visible in the optical in the AGN sample in this work, we believe that projection effects  causes the nuclear ionised gas to be often seen on top of the nucleus, and due to the low Av values, to appear  interlaced with the central filaments and lanes. 
 
 With the exception of M87, which is free of dust, dust and ionised gas are overlapping in NGC\,1052, NGC\,3081, NGC\,1566 and M58. It is notorious the case of M58 with a $7\, \rm{kpc}$ length dust filament and an equivalent counterpart in H$\alpha$ running North of nucleus (Fig.\,\ref{fig:m58}). The extinction in the filament is moderate, $A_V \sim 1.6\, \rm{mag}$ (Fig.\,\ref{fig:ext_1st}) insufficient to suppress H$\alpha$. In general, the extinction \av \ towards the centre in these galaxies is Av $<$ 2 mag whereas  in    the equivalent sample of type 2, \av \ reach factors   $2$--$3$ larger next to the centre (\citealt{Prieto2014}).

Only in those galaxies with the highest  extinction towards the centre, $A_V \sim 2$--$3\, \rm{mag}$, an spatial anti-correlation between the dust and the ionised gas is observed. {The best illustration is NGC\,1365 where  the [OIII] nuclear extended emission   is preferentially located at the  Eastern side of the nucleus, just where    Av is found the lowest,  $A_V \lesssim 1\, \rm{mag}$  (Fig.\,\ref{fig:ext_2nd}). No [OIII] is seen at the Western side due to an extended dust lane
 crossing  from North -East to South- West.   The extinction  in the lane is  $A_V \sim 2$--$3\, \rm{mag}$  (Fig.\,\ref{fig:ext_2nd}), sufficient to suppress  \oiii emission close to the centre and across the lane except for  a few dust-free patchy regions at the South and    at the sharp end of the lane, at its  Western side, where [OIII] re-appears.  (Fig.\,\ref{fig:ngc1365}). The dust map also unveils a number of star forming regions in the ring of NGC 1365 that were hidden by dust but get recovered in the K-band.They do not show [OIII] emission.}

\begin{figure}
	\includegraphics[width=0.5\textwidth]{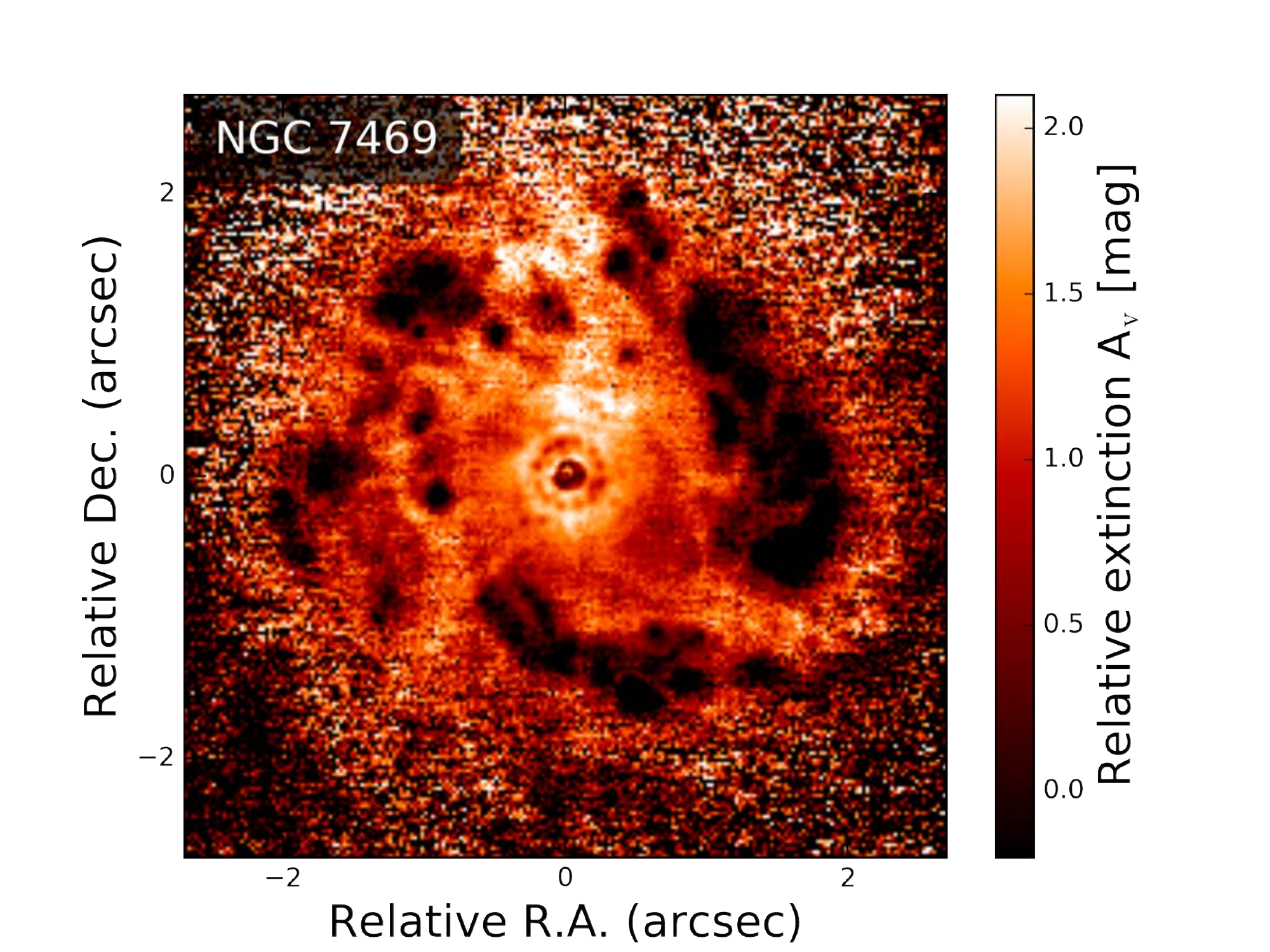}
	\caption{Extinction map of NGC\,7469.}\label{fig:ext_3rd}
\end{figure}

NGC\,1097 and NGC\,7469 have slightly lower extinction than NGC 1365, $A_V \lesssim 2\, \rm{mag}$ at the nucleus  surrounding, and in both, dust and gas tend to avoid each other particularly at those areas where the filament or lane gets optically thicker ((Fig.\,\ref{fig:ngc1097} and (Fig.\,\ref{fig:ngc7469}. In both, the ionised gas is associated with their respective circumnuclear star forming rings. At the centre, NGC\,1097 shows moderate anti-correlation between the nuclear dust spiral and \halpha (Fig.\,\ref{fig:ngc1097}, central and bottom panel); NGC\,7469 shows a relatively symmetric \oiii morphology at the nucleus, slightly extending East and South but sharply depressed towards North where a thick filament, $A_V \sim 2\, \rm{mag}$, crosses the star forming ring towards the nucleus (Fig.\,\ref{fig:ngc7469} and Fig.\,\ref{fig:ext_3rd}). 

{ To summarise, nuclear filaments and lanes, depending on their optically thickness and / or  their location with respect to the nucleus, can  give rise to the particular  gas morphologies  seen in type 2 and 1 alike. Collimated ionised gas is more frequent in type 2, which in is general understood in terms of the orientation of the torus with respect the line of sight. But as illustrated by previous works in these AGN type (\citealt{Prieto2014}; \citet{Mar2016}) the collimation may be a consequence of the location of dust filaments respect to the nucleus and of their optical thickness, both effects combine to extinguish totally or partially the  ionised gas. In the current type 1 -1.9 sample,   the filaments optical thickness is   insufficient to suppress the extended ionised gas in most cases, hence gas and dust interlace in the centre, and no collimation or preferential direction in the ionised gas is seen,  with  two exceptions:} NGC\,1365, caused by a  high Av dust lane passing next to the nucleus,  and NGC\,3081,  possibly caused  by  a patchy dust-free zone  North of  the nucleus.
A case aside is  NGC 1052, where the  gas collimation is possibly caused by the jet.

\subsection{ Dust filaments traced by molecular gas: streaming material to the centre}
The extinction maps in Figs 1 - 3 unveil the location, morphology and extension of obscuring material, presumably dust, in the centre of these galaxies. Filaments, resolved in nuclear spirals or in straight line to the centre, are traced up to tens of parsecs from the centre, suggesting they are channels transporting material to the nucleus \citep[e.g.][]{Prieto2005,Fathi2006,Espada2017}. Their morphology and volume densities --\,inferred from the extinction maps\,-- together with information of the velocity of the flow, can be used to estimate a range of inflow rates. Accordingly, we look for their counterpart in molecular tracers to infer on the kinematics. The extinction maps are deep reconstructions of most of the obscuring material in the galaxy centre. This material may be associated with different gas phases (\citealt{Mar2015}) and their counterpart in molecular gas  requires investigation of different line transitions levels.

The $300\, \rm{pc}$ length nuclear dust spiral and traced up to the central $5\, \rm{pc}$ radius in NGC\,1097 (Fig.\,\ref{fig:ext_1st} and \citealt{Prieto2005}) has been associated with warm-$H_2$ and HCN(4--3) molecular lines from which inflow rates less than $0.6\, \rm{M_\odot\,yr^{-1}}$ at $100\, \rm{pc}$ to  $0.2\, \rm{M_\odot\,yr^{-1}}$ at $40\, \rm{pc}$ from the centre are inferred (\citealt{Davies2009}; \citealt{Fathi2013}).

\begin{figure*}
	\includegraphics[width=0.3\textwidth]{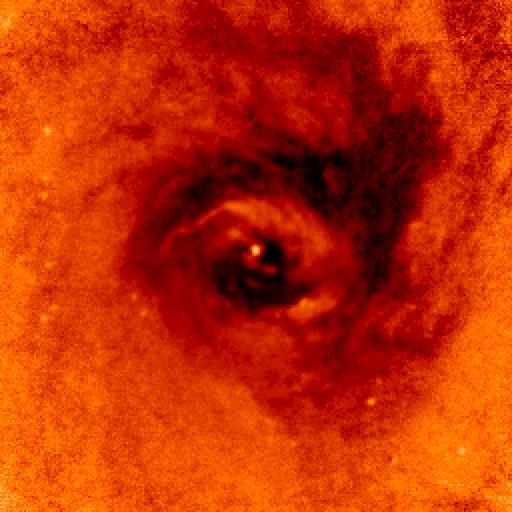}
	\includegraphics[width=0.3\textwidth]{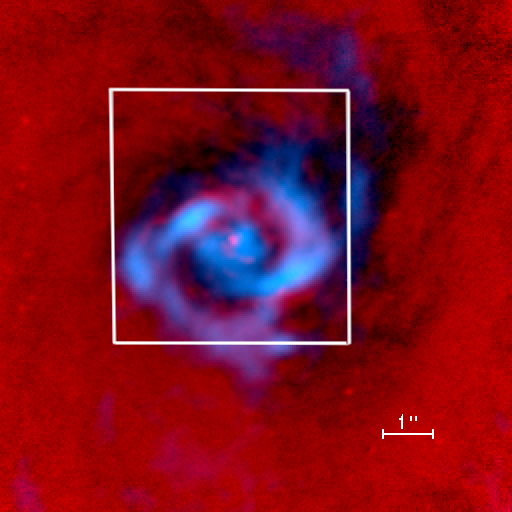}
	\includegraphics[width=0.3\textwidth]{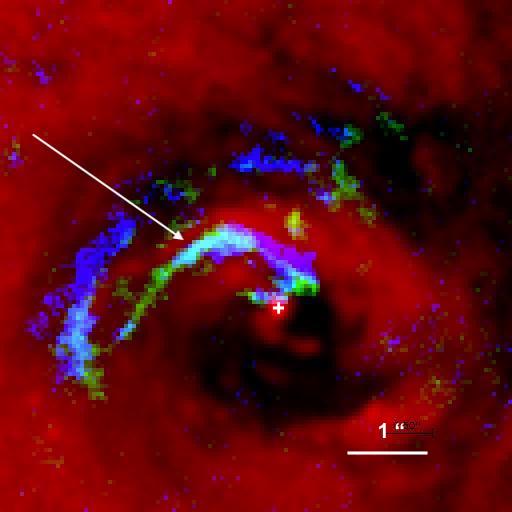}
	\caption{\textbf{Left:} NGC 1566 extinction map of Fig.\,\ref{fig:ext_2nd} but with color bar inverted, i.e., dark color means dusty regions. The nucleus is the white bright point-like source at the centre. \textbf{Centre:} Colour composition image of the extinction map (left panel) in red and the low resolution ALMA CO(3--2) map, beam $\sim 0\farcs30 \times 0\farcs36$ in blue. The nucleus is the pink color point-like source at the centre. \textbf{Right:} zoom of the central  half kpc - square in middle panel:  the extinction map is in red, 	
the CO(3--2) velocity channels in the range $20$--$80\, \rm{km\,s^{-1}}$, extracted from the higher angular resolution ALMA map, beam $60 \times 40\, \rm{mas}^{2}$, are shown on top. The low velocity channel shows in green, the high velocity one, in blue. This  velocity selective channel map isolates one of the nearest-to-the-centre dust filaments  (marked with an arrow, see text). A second dust filament at larger radii is also seen  tarced   in CO(3-2). The nucleus is marked with a cross. In all panels, North is up, East to the left.  }	\label{fig:alma}
\end{figure*}

NGC\,1566 is a  counterpart example of the NGC 1097 nuclear spiral  on a much larger   and denser scale  of filaments and lanes extending beyond 1 kpc (Fig.\,\ref{fig:ext_2nd}).  As in NGC 1097, a CO kinematic analysis by  \citet{Combes2019} and \citet{Slater2019}   evidence  inflow to the centre also   in NGC 1566, so we take this case to further illustrate the counterpart of  dust map in molecular gas and quantify an inflow rate. Within the central two arcsec radius, $\sim 200\, \rm{pc}$, a number of well defined, collimated dust filaments are seen circularising to the centre,  these filaments  can also be traced  in the high resolution ALMA CO(3--2) $345.8\, \rm{GHz}$ map obtained by  F. Combes (PI  2016.1.00296.S, beam size $\sim 0\farcs30 \times 0\farcs36$). The dust - gas  association in the central few hundred parsec can be seen in the colour composition image in Fig.\,\ref{fig:alma}. The figure shows a high contrasted, large-field of view dust map of the galaxy in red --\,from same image as that in Fig.\,\ref{fig:ext_2nd}\,-- with ALMA CO(3--2) gas on top in blue colour. The CO(3--2) map was  extracted from the ALMA pipeline-generated   beam-corrected data cube after continuum emission subtraction. To register this image to the dust  map,  ALMA line-free continuum emission maps at distinct frequencies were evaluated, all  showing a single point-like source at the centre of the field which was  assumed to be the nucleus. We used   this source as reference to register the ALMA CO(3--2) map to the IR-Optical-UV images, the  nucleus being at all these wavelengths  unambiguously identified. The CO(3--2) map presented here is a sum up of the velocity channels with  two sigma level signal in the line and  no background suppression applied to unveil all possible emission counterpart to the dust map also at larger radii from the centre (a cleaner background suppressed map is shown in Combes et al. 2019).
 
Fig.\,\ref{fig:alma}-central-panel isolates one of the longest, best colimated dust filaments in the higher resolution ALMA map, beam $60 \times 40\, \rm{mas^2}$. This map was extracted from the ALMA archive, and after reviewing each velocity channel  in comparison with the dust extinction map of Fig.\,\ref{fig:ext_1st}, it was possible to identify some of the dust filaments in the CO(3-2) map, which appear spatially resolved by their kinematics printed in the gas. For best illustration, Fig.\,\ref{fig:alma}-right-panel shows a selective integration map in  velocity channels, namely the blueshifted  velocity channels from $20\, \rm{km\,s^{-1}}$ to $80\, \rm{km\,s^{-1}}$, which show one of the most prominent dust filaments with counterpart in  CO(3--2) - the total integrated  velocity map   is  in \citealt{Combes2019}, their fig.\,20. This selective velocity map (Fig.\,\ref{fig:alma}-right-panel) unveils a  coherent structure in morphology, describing  a circularised trajectory  over circa  $150\, \rm{pc}$ length with  progressive decreasing pith angle  towards the centre. The velocity along the filament seems  increasing   to the centre.
 
We use this filament for getting an estimate of the inflow mass rate on the assumption that the gas in the inner few hundred parsecs is predominantly inflowing as inferred by the kinematic analysis in \citet{Combes2019} and \citet{Slater2019}. In the later work, the selected filament is outlined in the rotation-subtracted CO(2--1) map, albeit with low signal-to-noise, at the equivalent range of blueshifted velocities as those identified by us in Fig.\,\ref{fig:alma}-right-panel. From the published kinematic map by these authors, a  velocity of $ \sim 80\, \rm{km\,s^{-1}}$ is identified along this filament. For  velocity de-projection, an inclination angle to the line of sight of 28 degrees is assumed (e.g. \citealt{Garcia-gomez1991} and references therein).  { A maximum inclination of  90 degrees would translate into an inflow rate a factor two smaller.}
The volume density and filament width  are  estimated from the extinction map. The filament width, W, corresponds to a FWHM\,$\lesssim 200\, \rm{mas}$, or $20\, \rm{pc}$. The optical thickness in the extinction map increases towards the centre reaching $A_V \sim 1.5\, \rm{mag}$ within the central $30\, \rm{pc}$ radius (Fig.\,\ref{fig:ext_1st}). Assuming a dust to gas ratio of $2 \times 10^{21}\, \rm{cm^{-2}\,mag^{-1}}$ \citep{Savage-Mathis1979} and the filament width equal to its depth, the volume density in the central few tens of parsec is $n_{\rm e} \sim 50\, \rm{cm^{-3}}$. As the filaments are spatially separated, a filling factor in the filament equal to one is assumed. The inflow mass rate follows as:

\begin{equation}
\dot{M} \sim < 0.16\, \rm{M_\odot\,yr^{-1}} \times \left( \frac{n_e}{50\, \rm{cm^{-3}}} \right) \times \left(\frac{W}{20\, \rm{pc}} \right)^2 \times \left( \frac{V_{flow} \times \sin 28}{80\, \rm{km\,s^{-1}}} \right) \times 2
\end{equation}

The factor 2 accounts for a second much shorter filament  seen South-West the nucleus at the redshifted velocities in \cite{Combes2019}  kinematic map, presumably also approaching the centre. A filling factor a factor 10 smaller would imply an infall rate smaller by the same factor.
NGC\,1566 is a low Eddington Seyfert type 1.5: $L_{\rm bol}/L_{\rm edd} \sim 4 \times 10^{-3}$ ($L_{\rm bol}$ from Table\,\ref{tab:extinction}, BH mass $\sim 10^7\, \rm{M_\odot}$). {The estimated inflow rate, $ \sim < 0.2\, \rm{M_\odot\,yr^{-1}}$ is more than enough to support the mild activity of this nucleus, with  an Eddington  accretion rate  two order of magnitude lower}.   The inferred  inflow rate is in the range of those few reported for low Eddington sources: the LINER- Type 1 NGC\,1097,  $\dot{M} < 0.2\, \rm{M_\odot\,yr^{-1}}$ at $40\, \rm{pc}$ from the centre  (quoted above),  the Seyfert  2 NGC\,1386,  $\dot{M} < 0.4\, \rm{M_\odot\,yr^{-1}}$ within $70\, \rm{pc}$ from the centre (\citealt{Rodriguez-Ardila2017}), { and in both, these rates are orders of magnitude larger than  their Eddington accretion rates.} In comparison, the estimated inflow rate  for the higher, by an order of magnitude, Eddington accreting source NGC\,1068 (Lbol / Ledd $\sim 4 ~\times 10^{-2}$, Lbol  integrated from the parsec-scale SED in \citealt{Prieto2010}), is  an order of magnitude higher, a few $\rm{M_\odot\,yr^{-1}}$  at  $90\, \rm{pc}$ from the centre (\citealt{MullerSanchez2009}; \citealt{Tacconi1994}). { Accretion is fundamentally driven by the disc efficiency  to loose angular momentum, still the above few values hint  for higher inflows promoting higher accretion rates.}

\section{Conclusions}\label{sec:conclusions}

Parsecs scale dust maps of the central kpc of the nearest type 1--1.9 AGN unveil the ubiquitous presence of dust filaments and lanes spreading all over the central ten to hundred parsecs. The high-angular resolution of the maps permits us to resolve the filaments in long,  collimated structures with widths $< 10\, \rm{pc}$ FWHM. They can be traced from hundred- to kiloparsecs distance all the way to the centre. The way the filaments approach the centre is varied, either circularising and often forming a nuclear spiral, or getting straight to --\,or across\,-- the nucleus.
 
The morphology, ubiquity, and location, of these dust structures are the same as those seen in an equivalent sample of the nearest type 2 AGN (\citealt{Prieto2014}). The key difference is that in the type 2 AGN, the nuclear filaments happen to be optically thick enough to obscure the nucleus at the optical/UV wavelengths, whereas in type 1 and intermediate-type, the optical thickness at the nucleus is a factor 2--3 lower, hence the nucleus is always visible. In the type 2 cases, the nucleus becomes visible from $2\, \rm{\micron}$ onwards as the most outstanding point like source in the central kpc.  Astrometry to the $10\, \rm{mas}$ level shows their optical photometric peak to be always shifted with respect to the IR peak --\,the nucleus\,-- by few parsecs and to be associated typically with a star forming region, an ionised gas cloud or simply the stellar peak. In the present type 1 and intermediate AGN sample, the  central point-like source seen at $2\, \rm{\micron}$ falls within our precision at the optical peak location, as expected. 

Because of the filamentary morphology of the  dust, nuclear ionised gas, e.g H$\alpha$, and especially the shortest wavelength gas emission, as \oiii$\lambda 5007$, are expected to be seen in filament-free regions. This is indeed the case in the type 2 sample where gas and dust filaments show a sharp spatial anti-correlation. On the contrary,  { in the present sample, the ionised gas spreads all over the nuclear region in most cases regardless of the dust location. This overlap may be due to two reasons: 1) the low extinction of the filaments and lanes, typically $A_V < 2\, \rm{mag}$; 2)  because of 1) and the fact that the nucleus is  visible at all wavelengths, projection effects facilitate a cleaner view of the ionised gas on top the nucleus.}

{ The  extinctions derived from dust maps in this and \citealt{Prieto2014} works are in the range of few Av in  type 1, this work,  up to $\sim 6 - 8$  magnitudes in type 2, in the later sufficient to obscure the nucleus. Still, higher nuclear extinctions are   inferred in some type 2  from   mid-IR (broad-line detection and interferometry,  e.g. \citealt{Lutz2002}, \citealt{Burtscher2015} and X-rays (e.g \citealt{Ricci2015}). The  extinctions inferred from the present dust maps  strictly refer to  regions at a few parsec  from the AGN - this is any  nuclear color is contaminated by  the AGN emission. Yet,  as discussed in  \citealt{Prieto2014}; higher  extinctions are  foreseen at the AGN location itself due to two observational facts inferred from the dust maps: 1) the  increase in  optical depth along the filaments towards the centre, particularly  in type 2 (and further indicated  at subparsec scales  in e.g. the Circinus galaxy  (\citealt{Tristram2014}; \citealt{Mar2016}); 2) the filament nuclear spiral morphology   which suggests, and  confirmed in a number of cases (sect 3.3),   mass streaming  to the centre.
The arrival of filaments to the centre occurs in any direction, so the  frequently seen dust in the polar direction (cf. Introduction sect.) may be associated to this infall.}

Putting all together, a possible scenario arises in which nuclear obscuration in AGN, regardless of their type, may  be caused by  large, kpcs to  several hundred parsec,   filaments and lanes ubiquitous to centres of galaxies. Depending on the position of these filaments at the centre, their optical thickness and the brightness of the nucleus, a fully or partially obscured nucleus may arise.
 The scenario is in line with previous proposals put forward  by \citet{Malkan1998}  and \citealt{PoggeMartini2002}, on the basis  of \textit{HST} WFPC2  imaging  of a large sample of AGN. We shall add that the commonly observed central dust structures in their maps may arise at galaxy scales, and not necessarily  at the central parsecs. Also their frequency regarding AGN type is higher than anticipated by those authors as they appear  to be ubiquitous to  type 2 and 1 AGN alike. Galaxies, particularly of early type may often appear free of dust over their larger part of their body, as illustrated by many cases, particularly in type 1 AGN, in \citet{Malkan1998} \textit{HST} survey. However, when scrutinised with the high angular resolution and wide wavelength range dust maps used here, dust at their very centre is  the norm. Examples of this are NGC\,1052 (Fig.\,\ref{fig:ngc1052}, this work) or the type 2 MCG-5-23-016 \citep{Prieto2014}, exceptions would be e.g. M87 (Fig.\,\ref{fig:m87}), nuclear dust-free.  The large scale of these filaments and lanes naturally account for the often collimated  optical ionised gas structures seen in type 2 sources, this is not the case in type 1 due to the characteristic  low optical thickness  of the filaments in this case, which in turn it is line  with the less frequent gas collimation seen in  type  1.
 
 Filaments and lanes  are traced by equivalent structures in molecular  cold and warm gas in CO, HCN and  $H_2$. A single filament can be seen either getting straight- or spiralling- to the centre, suggesting  their role as streamers to channel material from  galactic scales to the centre.  Kinematic evidence is proving that  be the case in different degrees, from  a few tenth of solar masses per year in low Eddington sources to solar in the standard 1\% - 10\% Eddington cases. 
{ These large-scale dust structures may end up in the high extinctions inferred at parsec scale in type 2 AGN, and thought to be caused by a torus. The difference  with the present scenario is that obscuration arises naturally, by galactic-scale filaments and lanes ubiquitous to any galaxy converging to centre.  These structures may  fulfil the role of the putative parsecs-scale torus.}


Data availability: the data underlying this article will be shared on reasonable request to the corresponding author.

\section*{Acknowledgments}
MAP thanks the support  of the Excellence Cluster ORIGINS which is funded by the Deutsche Forschungsgemeinschaft (DFG, German Research Foundation) under Germany Excellence Strategy - EXC-2094 - 390783311. JN acknowledges financial support by the project Evolution of Galaxies, of reference AYA2014-58861-C3-1-P and AYA2017-88007-C3-1-P. Substantial part of this work formed part of the Master thesis of JN. JAFO acknowledges financial support by the Agenzia Spaziale Italiana (ASI) under the research contract 2018-31-HH.0; MM to  the Ramon y Cajal fellowship RYC2019-027670-I. F. Combes kindly provided us with ALMA data of NGC 1566. This research  made use of the NASA/IPAC Infrared Science Archive, which is operated by the Jet Propulsion Laboratory, California Institute of Technology, under contract with the National Aeronautics and Space Administration.

\bibliographystyle{mnras}
\bibliography{mystrings,biblio}

\appendix

\begin{figure*} 
\section{Figures}\label{sec:figures}
 \centering\includegraphics[width=0.9\textwidth]{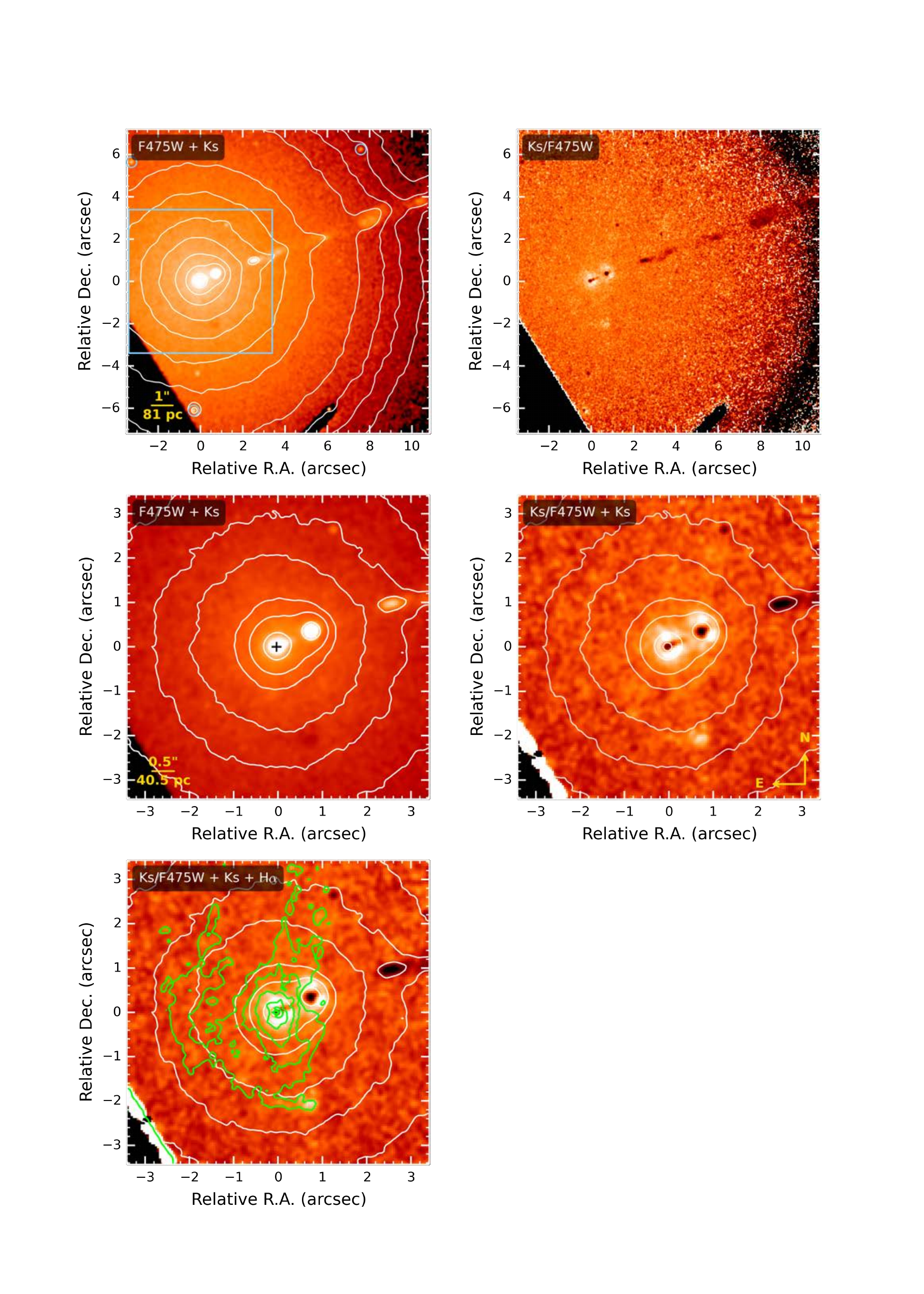}\vspace{-20pt}
 \caption{M87. \textbf{Top left}: \textit{HST} / F475W image with VLT-NaCo / \textit{Ks} band contours in white. The FoV is $14'' \times 14''$. The blue square marks the zoomed area  in middle and bottom panels. Blue circles are the point-like sources used for images registration. The cross in middle and bottom panels marks the nucleus position and  error. \textbf{Top right:} \textit{Ks} / F475W  ratio or  dust map, {nucleus and  jet shows dark because of  their intrinsic bluer colors.} \textbf{Middle left:} F475W image with  \textit{Ks}-band contours in white.\textbf{Middle right:} \textit{Ks} / F475W - dust map - with the \textit{Ks}-band in contours in white. \textbf{Bottom left:} As the former with H$\alpha$ in green contours.}\label{fig:m87}
\end{figure*}

\begin{figure*} 
 \includegraphics[width=0.9\textwidth]{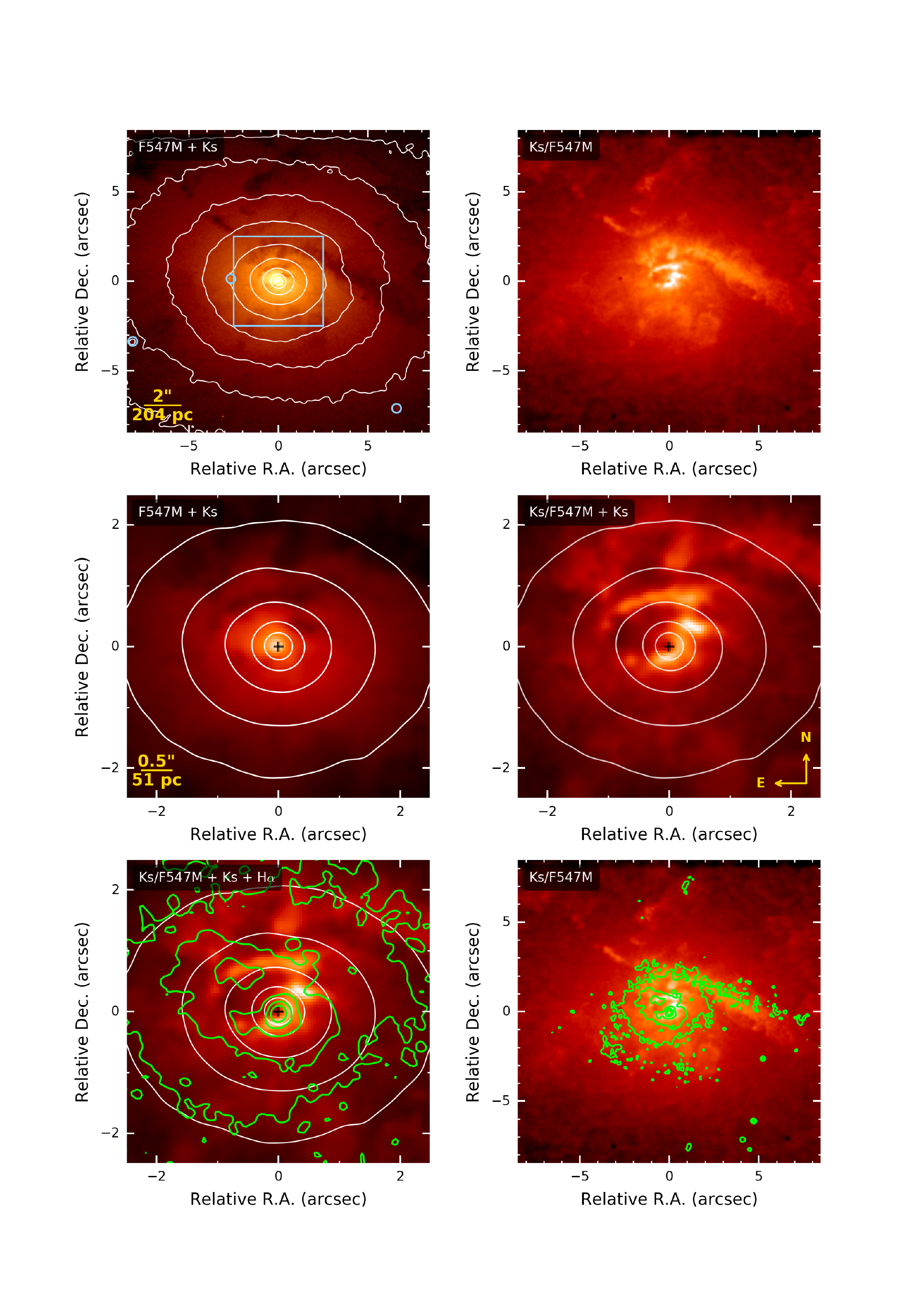}
\vspace{-20pt}
 \caption{M58. Marks in the images and contour colors follow same convention as in A1.  Optical comparison image in this case is  \textit{HST} / F547M. \textbf{Top left}: \textit{HST} / F547M image with VLT-NaCo / \textit{Ks} band contours. The FoV is $17'' \times 17''$.   \textbf{Top right:}  Dust map \textit{Ks} / F547M. \textbf{Middle left:} Zoomed area: F547M image with  \textit{Ks}-band contours. \textbf{Middle right:} Dust map with  \textit{Ks} band in contours. \textbf{Bottom left:} Same as former with H$\alpha$-in green.  \textbf{Bottom right:} Dust map with H$\alpha$ full FoV.}\label{fig:m58}
\end{figure*}
\begin{figure*} 
 \includegraphics[width=0.9\textwidth]{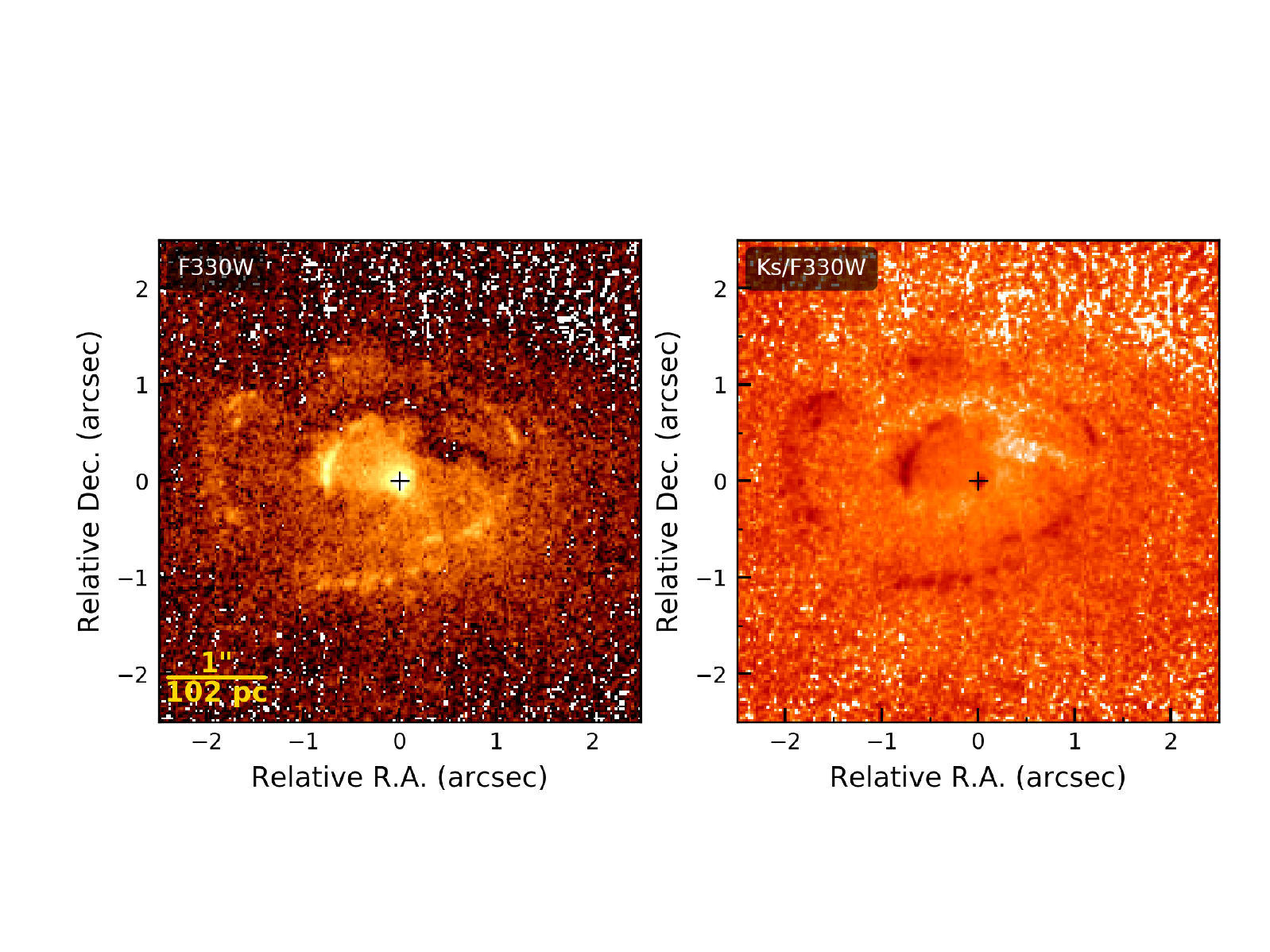}
\vspace{-20pt}
 \caption{M58. \textbf{Left}:  \textit{HST} / F330M image showing the UV nucleus, circumnuclear star forming ring, and the two main inner filaments --\,in dark colour\,-- enshrouding the nucleus. \textbf{Right}: Dust map VLT-NaCo \textit{Ks} /  \textit{HST} F330W.  The dust filaments show now  bright as  UV emission is highly depressed, the nucleus and stellar ring show darker because of their high UV emission. The cross marks the nucleus position and its error.}\label{fig:m58-uv}
\end{figure*}

\begin{figure*} 
\includegraphics[width=0.9\textwidth]{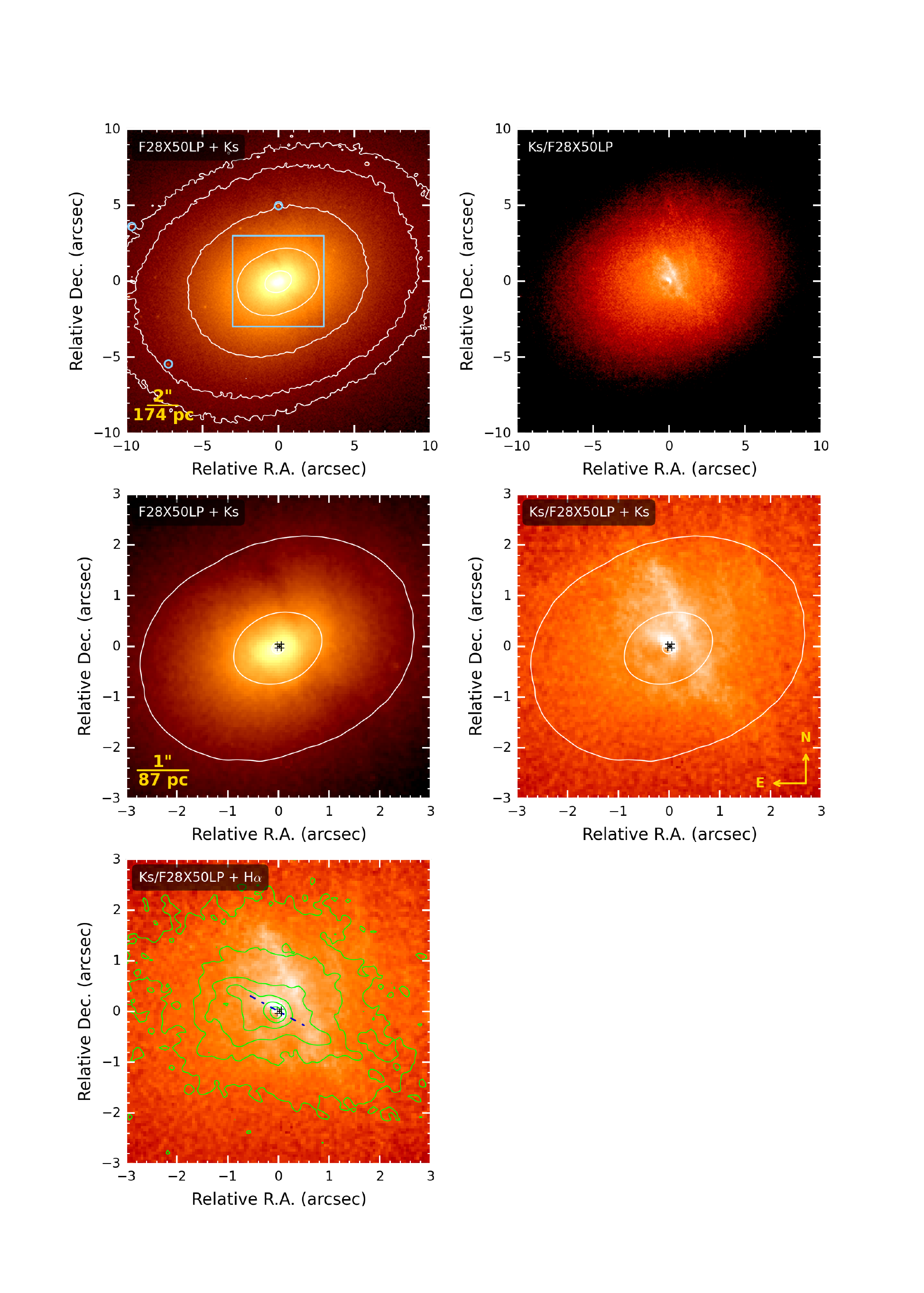}\vspace{-20pt}
 \caption{NGC\,1052.  Marks in the images and contour colors follow same convention as in A1. Reference HST image used in this case is \textit{HST}/F28X50LP.  \textbf{Top left}: \textit{HST}/F28X50LP image and VLT-NaCo / \textit{Ks} in contours. The FoV is $20'' \times 20''$.  \textbf{Top right:} \textit{Ks} / F28X50LP ratio or  dust map. \textbf{Middle left:} Zoomed area \textit{HST}/F28X50LP image and \textit{Ks}-band in contours.  \textbf{Middle right:} Dust map with  \textit{Ks}-band in contours. \textbf{Bottom left:} As previous but with  H$\alpha$ in green contours.  The dashed line indicates the parsec-scale twin-jet direction (\citealt{Kadler2004}).}\label{fig:ngc1052}
\end{figure*}

\begin{figure*} 
\includegraphics[width=0.9\textwidth]{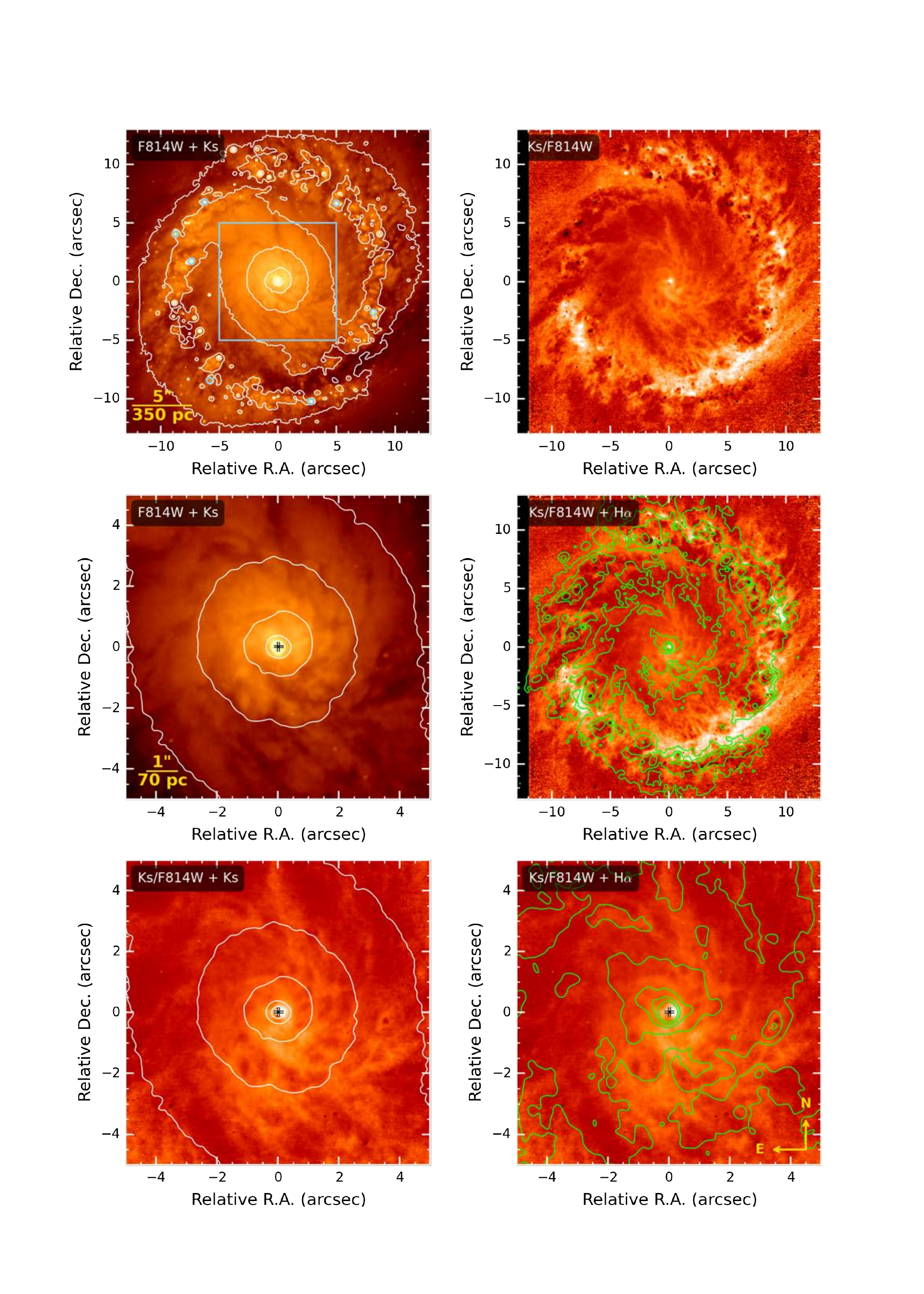}\vspace{-20pt}
 \caption{NGC\,1097. Marks in the images and contour colors follow same convention as in A1. Reference HST image used in this case is \textit{HST }/ F814W. \textbf{Top left}: \textit{HST} / F814W image with VLT-NaCo / \textit{Ks}-band in contours. The FoV is $26'' \times 26''$. \textbf{Top right:} Dust map \textit{Ks} / F814W. \textbf{Middle right:}  Dust map with H$\alpha$ in green. \textbf{Middle left:} Zoomed area \textit{HST} / F814W image with  \textit{Ks}-band in contours.  \textbf{Bottom left:} Dust map with  \textit{Ks}-band in contours. \textbf{Bottom right:} Dust map with H$\alpha$  in contours.}\label{fig:ngc1097}
\end{figure*}

\begin{figure*} 
\includegraphics[width=0.9\textwidth]{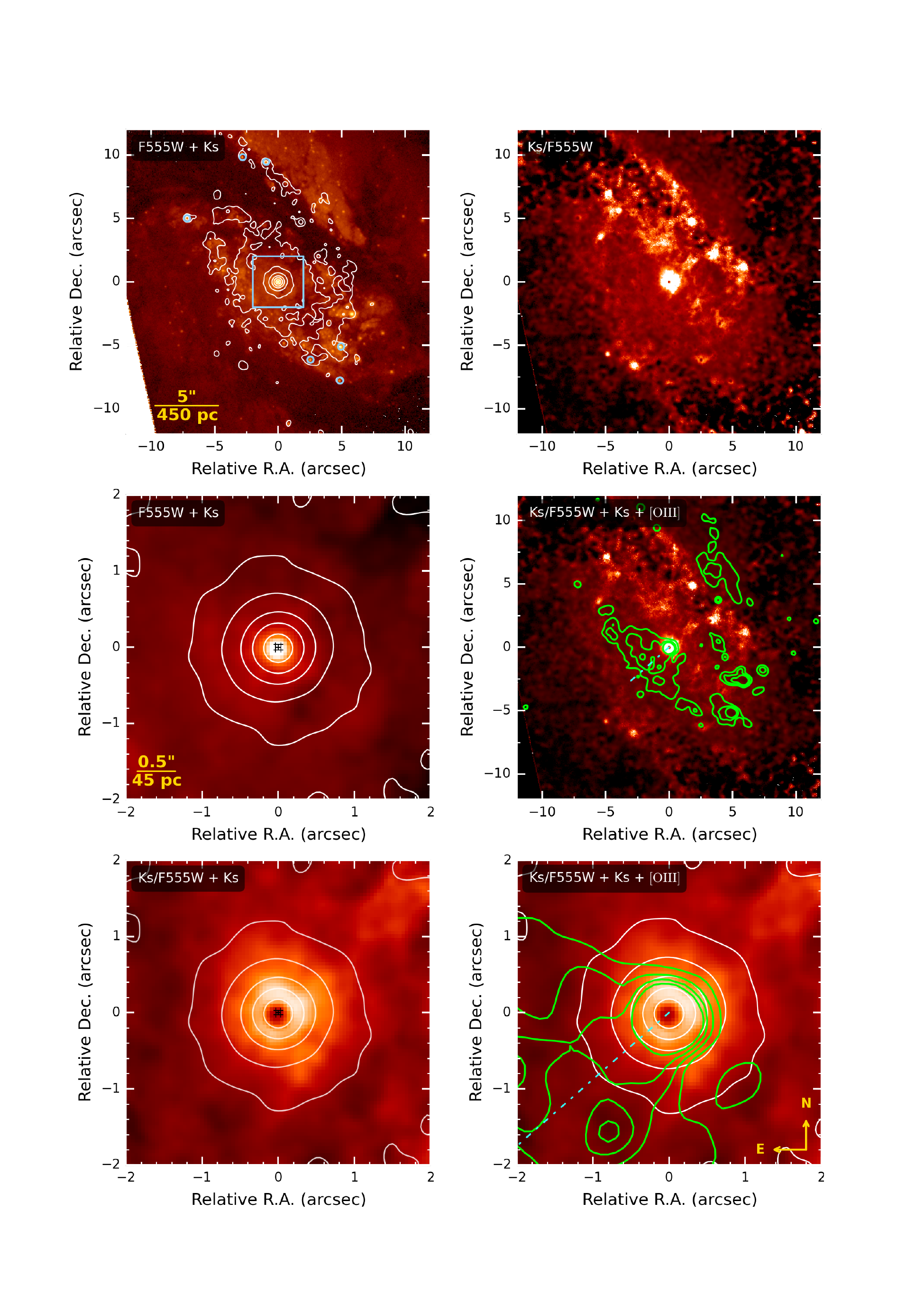}\vspace{-20pt}
 \caption{NGC\,1365. Marks in the images and contour colors follow same convention as in A1. Reference HST image used in this case is \textit{HST }/ F555W. \textbf{Top left}: \textit{HST} / F555W image with VLT-NaCo / \textit{Ks}-band in contours. The FoV is $24'' \times 24''$. \textbf{Top right:} Dust map \textit{Ks} / F555W. \textbf{Middle right:} Dust map with [\textsc{Oiii}] in contours. \textbf{Middle left:} Zoomed area F555W image with the \textit{Ks}-band in contours.  \textbf{Bottom left:} Dust map with Ks-band in contours.  \textbf{Bottom right:} As previous but added [\textsc{Oiii}] in green contours. The dashed line shows  the radio jet direction  (\citealt{Sandqvist1995})}\label{fig:ngc1365}
\end{figure*}

\begin{figure*} 
\includegraphics[width=0.9\textwidth]{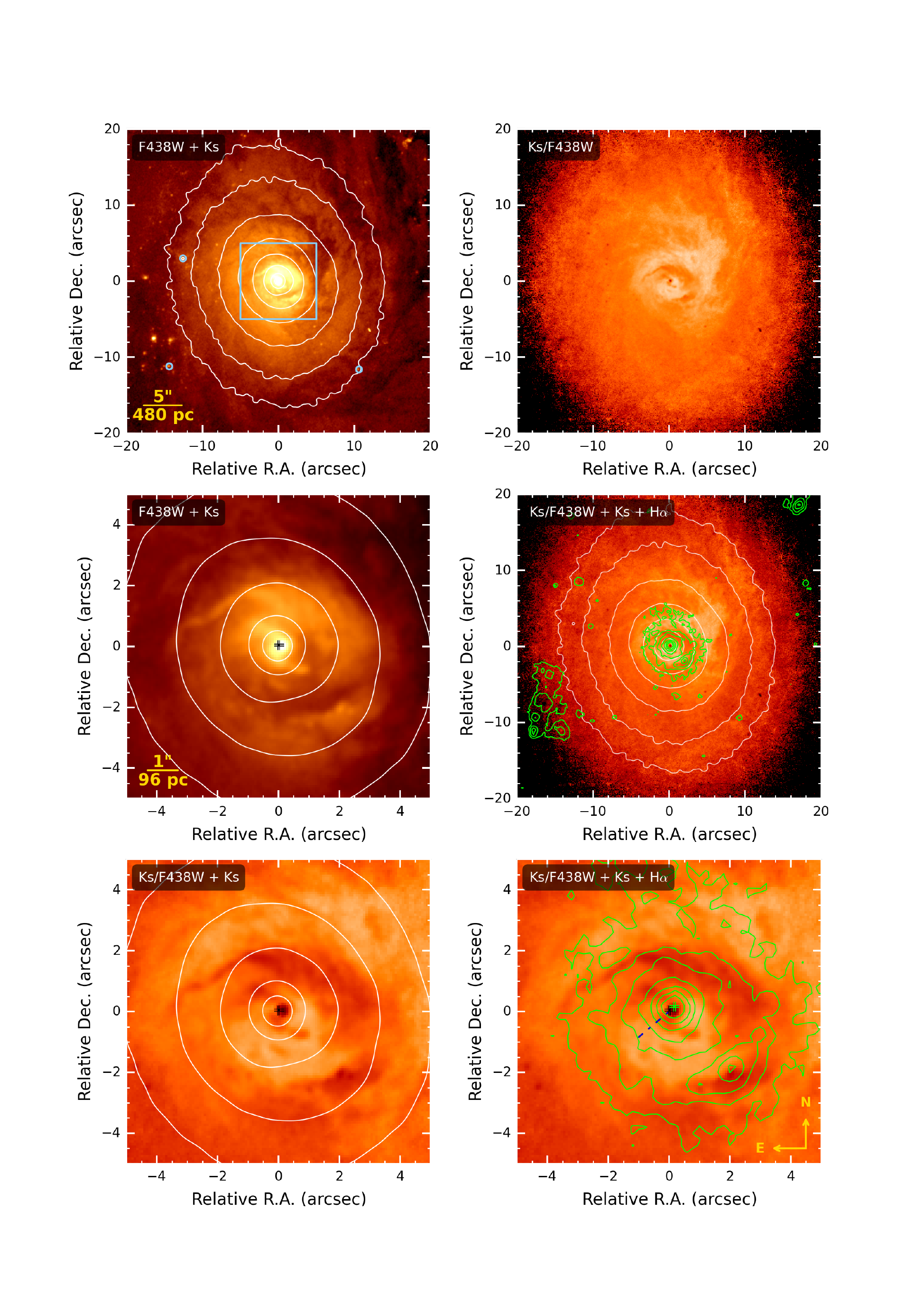}\vspace{-20pt}
 \caption{NGC\,1566. Marks in the images and contour colors follow same convention as in A1. Reference HST image used in this case is \textit{HST }/ F438W.
 \textbf{Top left}: \textit{HST} / F438W image with VLT-NaCo / \textit{Ks}-band in contours. The FoV is $40'' \times 40''$.  \textbf{Top right:} \textit{Ks} / F438W ratio or dust map. \textbf{Middle right:} Dust map with \textit{Ks}-band in white-,  H$\alpha$ in green-contours. \textbf{Middle left:} Zoomed area: F438W image with the \textit{Ks}-band in contours.  \textbf{Bottom left:} Dust map with Ks-band in contours.  \textbf{Bottom right:} Dust map with H$\alpha$ in contours. The H$\alpha$ peak is shown with  a green cross at the centre.}\label{fig:ngc1566}
\end{figure*}

\begin{figure*} 
\includegraphics[width=0.9\textwidth]{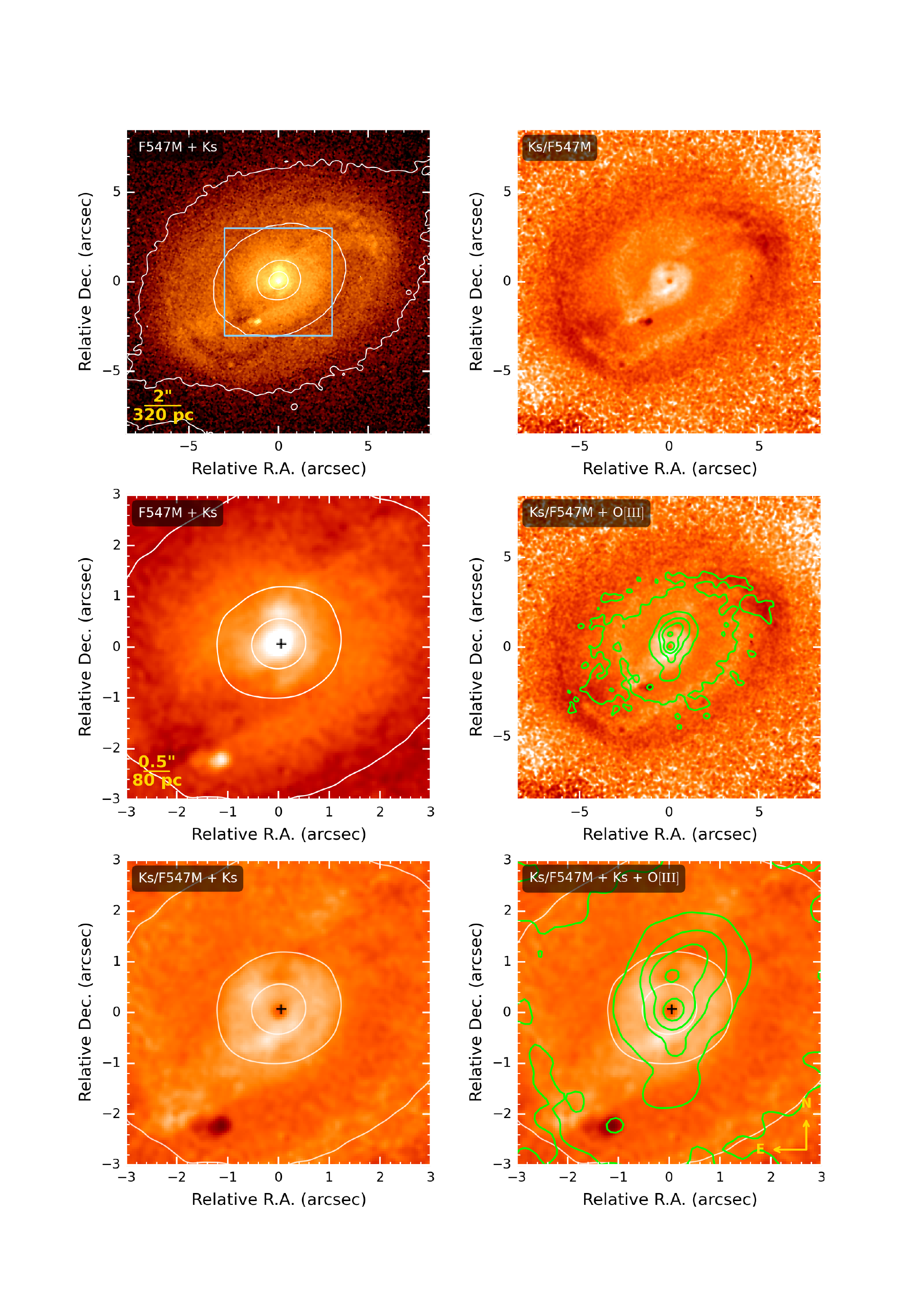}\vspace{-20pt}
 \caption{NGC\,3081. Marks in the images and contour colors follow same convention as in A1. Reference HST image used in this case is \textit{HST }/ F547M.  \textbf{Top left}: \textit{HST} / F547M image with VLT-NaCo / \textit{Ks}-band in contours. The FoV is $17'' \times 17''$.  \textbf{Top right:} Dust map \textit{Ks} / F547M.  \textbf{Middle right:} Dust map with [\textsc{Oiii}] in contours  \textbf{Middle left:} Zoomed area: F547M image with the \textit{Ks}-band in contours. \textbf{Bottom left:} Dust map with the \textit{Ks}-band in contours. \textbf{Bottom right:} As previous but added  [\textsc{Oiii}] in green contours.}\label{fig:ngc3081}
\end{figure*}

\begin{figure*} 
\includegraphics[width=0.9\textwidth]{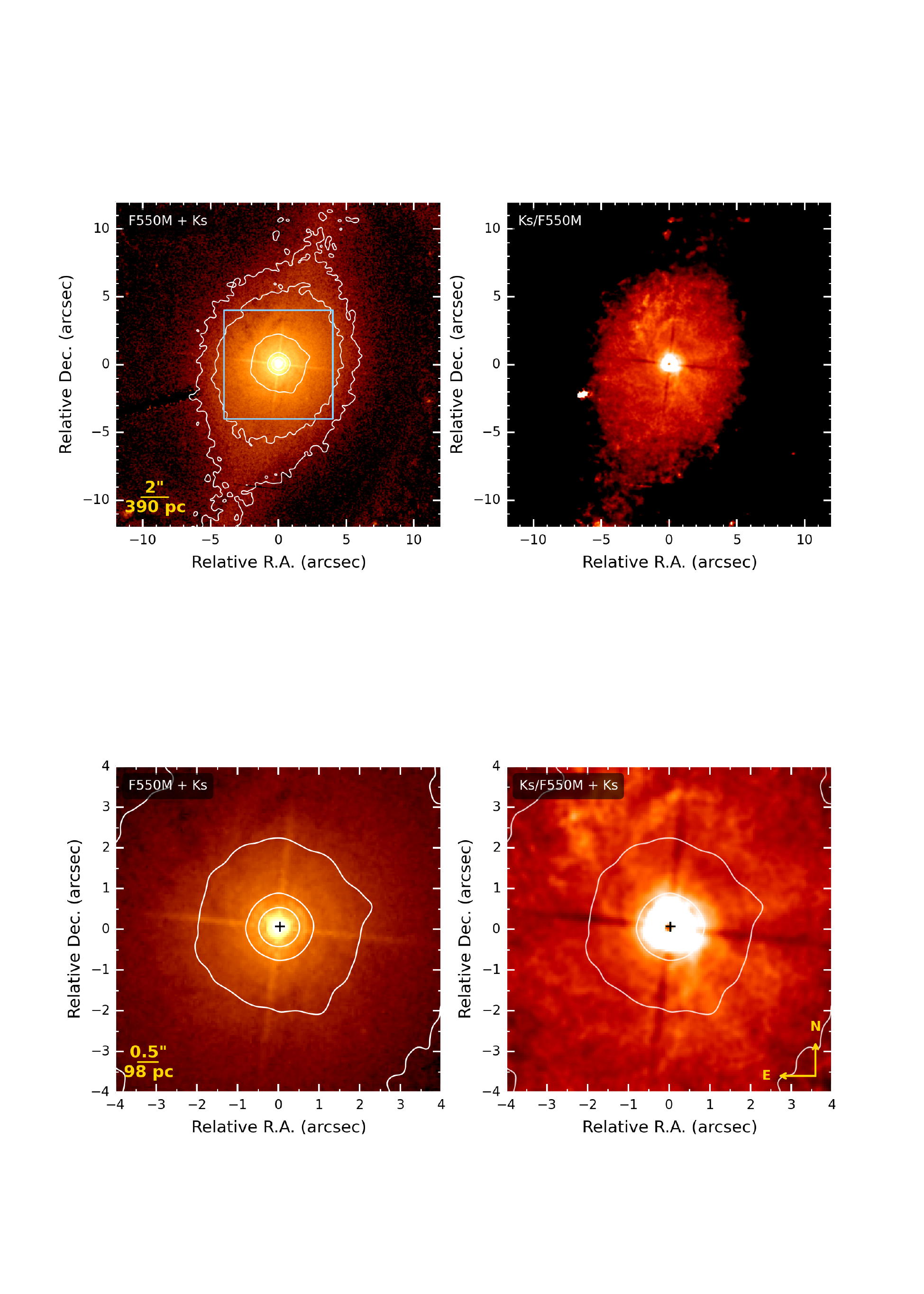}\vspace{-20pt}
 \caption{NGC\,3783. Marks in the images and contour colors follow same convention as in A1. Reference HST image used in this case is \textit{HST }/ F550M. \textbf{Top left}: \textit{HST} / F550M image with VLT-NaCo / \textit{Ks}-band in contours. The FoV is $24'' \times 24''$. \textbf{Top right:} Dust map \textit{Ks} / F550M. \textbf{Bottom left:} Zoomed area: F550M image with the \textit{Ks}-band in contours. \textbf{Bottom right:} Dust map with the \textit{Ks}-band in contours.}\label{fig:ngc3783}
\end{figure*}

\begin{figure*} 
\includegraphics[width=0.9\textwidth]{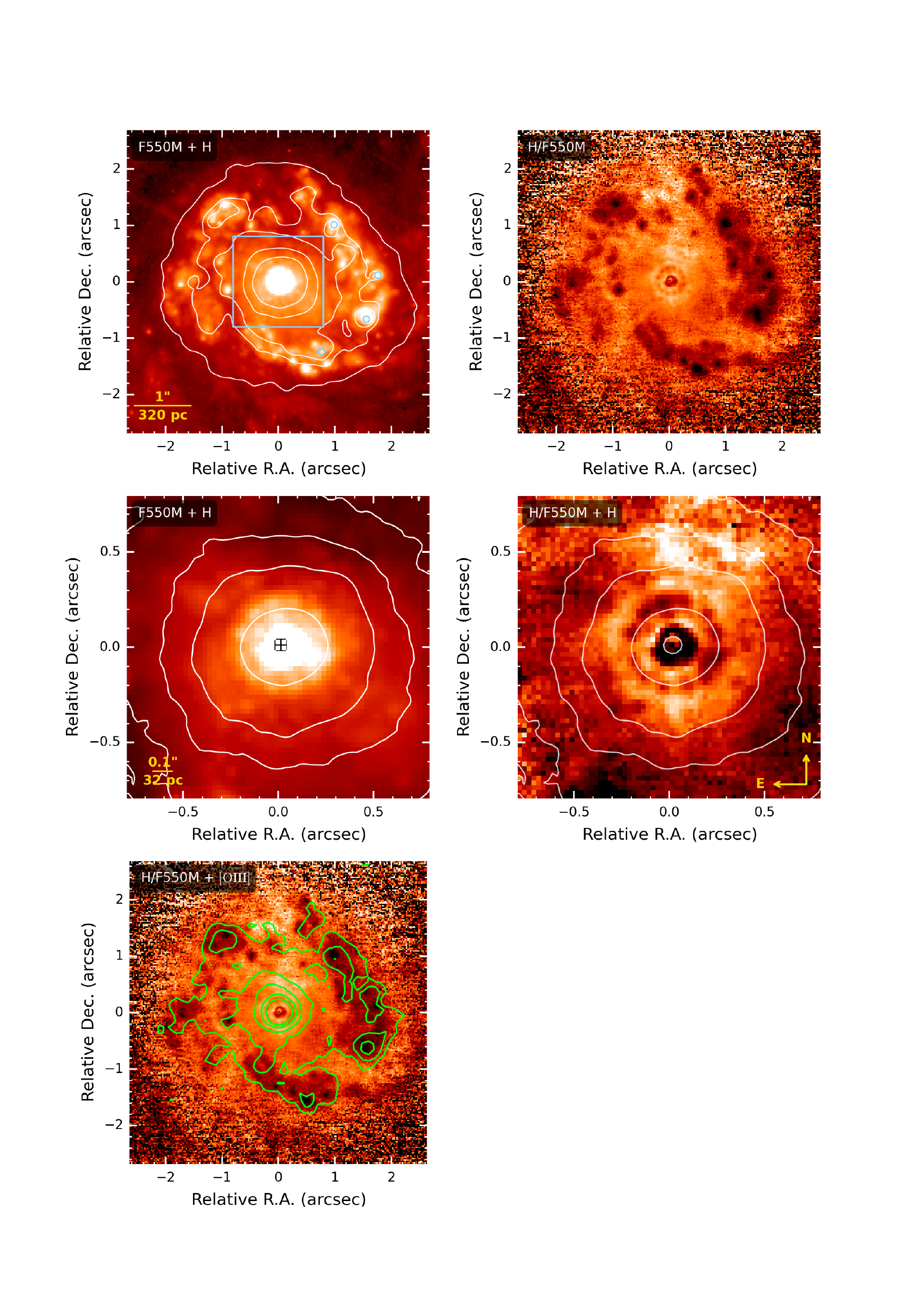}\vspace{-20pt}
 \caption{NGC\,7469.  Marks in the images and contour colors follow same convention as in A1. Reference HST image used in this case is \textit{HST} / F550M. \textbf{Top left}: \textit{HST} / F550M image with the \textit{HST} / \textit{H}-band in contours. The FoV is $5.4'' \times 5.4''$. \textbf{Top right:} Dust map \textit{H}-band / F550M. \textbf{Middle left:} F550M image with the \textit{H}-band in contours. \textbf{Middle right:} Zoomed area: dust map with the \textit{H}-band in contours. \textbf{Bottom left:} Dust map with [\textsc{Oiii}] in contours.}\label{fig:ngc7469}
\end{figure*}



\bsp	
\label{lastpage}
\end{document}